\documentclass[12pt]{iopart}
\usepackage{iopams}  

\usepackage{cite}
\usepackage{verbatim}
\usepackage{bm} 
\usepackage{bbm} 

\usepackage{graphicx}        

\usepackage{pstricks,pst-grad,pst-plot}

\newtheorem{theorem}{Theorem}[section]
\newtheorem{corollary}{Corollary}[theorem]

\usepackage{amssymb}

\begin{document}

\title[Bifurcations in synergistic epidemics]{Bifurcations in synergistic epidemics on random regular graphs}

\author{S.N. Taraskin}

\address{Dept. of Chemistry, University of Cambridge, Cambridge CB2 1EW, UK}
\address{St. Catharine's College, University of Cambridge, Cambridge CB2 1RL, UK}
\ead{snt1000@cam.ac.uk}

\author{F.J. P{\'e}rez-Reche}
\address{Institute of Complex Systems and Mathematical Biology, SUPA, School of Natural and Computing Sciences, University of Aberdeen, Aberdeen AB24 3UE, UK}
\ead{fperez-reche@abdn.ac.uk}
\vspace{10pt}
\begin{indented}
\item[]October 2018
\end{indented}

\begin{abstract}
The role of cooperative effects (i.e. synergy) in transmission of infection is investigated analytically and numerically for epidemics following the rules of Susceptible-Infected-Susceptible (SIS) model defined on random regular graphs. Non-linear dynamics are shown to lead to bifurcation diagrams for such spreading phenomena exhibiting three distinct regimes: non-active, active and bi-stable. The dependence of bifurcation loci on node degree is studied and interesting effects are found that contrast with the behaviour expected for non-synergistic epidemics.
\end{abstract}

%
\vspace{2pc}

\section{Introduction}
\label{sec:Introduction}

Epidemics in networks is an important and interesting topic attracting a lot of research activity~\cite{Pastor-Satorras_15:review,Barrat_08:book,Dorogovtsev_2008:RevModPhys,Porter-Gleeson_Book2016}.
Recently, significant attention was paid to discontinuous phase transitions observed for spreading phenomena in complex networks~\cite{Dodds_2004:PRL,Gross_2006:PRL,Marceau_2010:PRE,Baxter_2010:PRE,Gleeson_2013:PRX,Gomez-Gardenes_Lotero_Taraskin_FJPR2015,Chae_2015:New_J_Phys,Liu_ChengLai_PRE2017,Bottcher_2017:Sci_Rep,Bottcher_2017:PRL,Min_2018:Sci_Rep}.
Discontinuous phase transitions are typically exhibited by the so-called threshold models ~\cite{Pastor-Satorras_15:review}, such as the second Schl{\"o}gl's model for autocatalytic reactions~\cite{Schlogl_1972}, quadratic contact process~\cite{Liu_2007:PRL}, the Watts model~\cite{Watts_2002:PNAS,Majdandzic_2014:NaturePhys} and  generalised contact process~\cite{Bottcher_2017:Sci_Rep,Bottcher_2017:PRL}.
In these models, the nodes can be in different states which evolve according to certain dynamical rules.
Within threshold models, the state of nodes changes when a certain threshold is achieved, e.g. a node can change its state when the number of its neighbours exceeds some threshold.
Bi-stability regions for concentration of infected nodes characteristic of discontinuous transitions  were also found for an SIS model with adaptive network topology in which susceptible nodes try to avoid contacts with infected nodes by breaking the links with them and then rewiring these links to other randomly chosen nodes~\cite{Gross_2006:PRL,Marceau_2010:PRE,Guo_2013:PRE}.
Bootstrap percolation on uncorrelated complex networks~\cite{Baxter_2010:PRE} provides one more example of discontinuous behaviour. 
This is a two-state  model describing active and non-active nodes.
A node becomes active and remains in such state if $k$ of its neighbours are active. The size of the giant active component can exhibit a discontinuous transition when the initial concentration of active nodes is used as a control parameter.

The synergy model introduced in Ref.~\cite{Perez_Reche_2011:PRL} describes non-linear co-operative effects in communication between nearest neighbours in a network which can lead to discontinuous phase transitions in common epidemic models~\cite{Gomez-Gardenes_Lotero_Taraskin_FJPR2015}.
The key ingredient of the synergy model is that the transmission of infection is described by means of continuous (in contrast to discrete for the threshold models) functions  of  discrete variables such as the number of nearest neighbours in a certain state.
Motivation for including synergy effects into the model comes from the experimental observations for e.g.  soil-born epidemics~\cite{Ludlam_2011:Interface} and spreading phenomena in social networks~\cite{Centola_2010:Science,Juul_2018_synergy:Chaos}. 

In this paper, we discuss the role of cooperative (synergistic)  effects in transmission of infection and demonstrate how these effects can lead to  discontinuous transitions.
The analysis is undertaken for the SIS process spreading on random k-regular  graphs.
The aim of our analysis is two-fold and consists of developing a  minimal and sufficient analytical framework (single-site mean-field) capturing all significant effects and suggesting a numerical procedure for supporting analytical findings.
The description is presented in terms of bifurcation theory~\cite{Strogatz_NonlinearBook}  which naturally suits our aims.

\section{The model}
\label{sec:Contagion}

\subsection{The rate equation}

We propose a model for synergistic SIS spread on a network of $N$ nodes whose connectivity defines a random $k$-regular graph in which  each node is randomly connected to $k$ different nodes. 
The state of each node $i$ at time $t$ is characterised by a state variable $\sigma_i(t)$ which can be either $\sigma_i(t)=0$ or $\sigma_i(t)=1$ corresponding to susceptible (S) or infected (I) state, respectively. In a time interval $\delta t$, nodes can change their state according to the following dynamical rules. If node $i$ is in the state I (i.e. $\sigma_i(t)=1$) at time step $t$ then it can change its state to S at time step $t+\delta t$ (i.e. $\sigma_i(t+\delta t)=0$) with probability $\mu \delta t$.
Here,  the parameter $\mu$ is the recovery rate which is assumed to be independent  of the states of other nodes in the network and is the same for all nodes.
If node $i$ is in the state S ($\sigma_i(t)=0$) then it can go to the state I at step $t+\delta t$ due to infection transferred from  its 
$n_i(t)=\sum_{m=1}^{k}\sigma_{i_m}(t) \equiv \sum_{m=1}^{k}\sigma_{m}(t)$ infected neighbours where the node number $i_m$ of neighbour $m$ for brevity is replaced by just its nearest-neighbour index $m$.
This occurs with probability $\Lambda_{n_i(t)} \delta t$, where $\Lambda_{n_i(t)}$ is the total transmission rate.
The infection can be transmitted to node $i$ independently by any of its infected neighbouring nodes, $\{ i_j | j=1,\ldots,n_i(t)\}$, with probability $\lambda_{i_j i}\delta t$, where $\lambda_{i_j i}$ is the individual transmission rate of infection from neigbouring infected node $i_j$ to node $i$. For brevity, the transmission rate will be denoted as $\lambda_{ij} \equiv \lambda_{i_j i}$ in the following, i.e. the $j$-th infected neighbour of node $i$ is numbered by the index $j$.
For independent transmissions in a particular configuration for node $i$ surrounded by $n_i(t)$ infected nodes, the total transmission rate is related to individual transmission rates as
\begin{equation}
  \Lambda_{n_i(t)} \delta t=1-\prod_{j}^{n_i(t)}(1-\lambda_{ji} \delta t)~.
  \label{eq:Lambda-definition}
  \end{equation}

In a standard formulation~\cite{Pastor-Satorras_15:review,Barrat_08:book} which ignores cooperative effects, the individual transmission rates are assumed to be constant values, $\lambda_{ji}=\lambda$, which do not depend on the state of neighbours of node $i$, so that the total transmission rate equals $  \Lambda_{n_i(t)} =\left[1-(1-\lambda \delta t)^{n_i(t)}\right]/\delta t$. 
In case of the synergistic transmission, the individual transmission rates do depend on the neighbourhood of node $i$ and below we consider a case when individual transmission rates depend on the number of infected neighbours of node $i$, i.e. $\lambda_{ji}=\lambda_{n_i(t)}$, but do not depend on the properties of neighbour $j$ such as its degree~\cite{Perez_Reche_2011:PRL,Taraskin-PerezReche_PRE2013_Synergy,Gomez-Gardenes_Lotero_Taraskin_FJPR2015}. 

In order to make the definition~\eref{eq:Lambda-definition} of the total transmission rate for synergistic transmission clearer, let us consider a simple example for a particular local configuration $C$ in a $3$-regular graph.
  Assume that $C$ at time $t$ consists of a central uninfected node $i=0$, i.e. $\sigma_0(t)=0$, two infected neighbouring nodes $i_1 = 1$ ($j=1$) and $i_2 = 2$ ($j=2$), i.e. $\sigma_1(t)=1$ and $\sigma_2(t)=1$, and one uninfected neighbouring node $i_3 = 3$ with 
  $\sigma_3(t)=0$, or symbolically $C= C_2=0_0\cap 1_1\cap 1_2\cap 0_3$ (the subscript in $C_2$ indicates that there are two infected neighbours).
  Let $T_0$ be the event that the central node changes its state at time $t+\delta t$ through transmission of infection from the infected neighbours.
    The infection can be transmitted  to node $0$ by means of several independent (by assumption) events, i.e.  $T_0=(T_{10}\cap \overline{T}_{20})\cup (T_{20}\cap \overline{T}_{10})\cup (T_{10}\cap {T}_{20})$, where $T_{j 0}$ ($\overline{T}_{j 0}$) is the event of transmission (non-transmission) of infection from infected neighbour $j$ to node $0$. Alternatively, infection might not be transmitted and this corresponds to the event  
  $ \overline{T}_{0}= \overline{T}_{10}\cap  \overline{T}_{20}$.
  For a given configuration $C_2$, the probability for node $0$ to be infected is
  $P(T_0|C_2)=1-P(\overline{T}_0|C_2)= 1-P(\overline{T}_{10}\cap  \overline{T}_{20}|C_2)=1-P(\overline{T}_{10}|C_2)P(\overline{T}_{20}|C_2) =
  1-(1-P(T_{10}|C_2))(1-P(T_{20}|C_2))$, where $P(T_{j 0}|C_2)$ is the probability of transmission of infection from neighbour $j$ to node $0$ given configuration $C_2$.
  This conditional probability $P(T_{j0}|C_2)$ is defined in terms of individual transmission rates differently for synergistic and non-synergistic transmission.
  For non-synergistic transmission, $P(T_{j0}|C_2)=P(T_{j0}|0_0\cap 1_{j})=\lambda \delta t $, i.e. it is a non-zero constant value for all recipient-donor pairs in the network if recipient $0$ is in S-state  and donor $j$ is in I-state.
  In particular, the non-synergistic transmission probability  $P(T_{j0}|C_2)$ does not depend on the state of all other neighbours of node $0$ except node $j$ and thus $P(T_0|C_2)=1-(1-\lambda\delta t)^2$. 
  In contrast, for synergistic transmission, the value of  $P(T_{j0}|C_2)$ depends on the number of infected nodes in $C_2$, i.e.  $P(T_{j0}|C_2)=\lambda_2 \delta t $ ($j=1,2$), and thus  $P(T_0|C_2)=1-(1-\lambda_2\delta t)^2$ where index $2$ in the transmission rate refers to the two infected nodes in $C_2$.
  For configuration $C_n$ with $n$ infected neighbours surrounding a non-infected central node $0$, the synergistic transmission probability is
  $P(T_0|C_n) \equiv \Lambda_n\delta t = 1-(1-\lambda_n\delta t)^n$ while the non-synergistic is  $P(T_0|C_n) \equiv \Lambda_n\delta t = 1-(1-\lambda\delta t)^n$. 
  The two types of transmission are equivalent if $\lambda_n$ does not depend on number of infected neighbours of the recipient (susceptible central node), i.e. $\lambda_n=\lambda$. 

Under these dynamical rules, the change of the probability $P(\sigma_i(t)=1)\equiv p_i(t)$ per unit time $\delta t$ obeys the following equation:
\begin{equation}
  \frac{\delta p_i(t)}{\delta t} \equiv \frac{p_i(t+\delta t)- p_i(t)}{\delta t}= R(p_i,\{ \sigma_j(t)\})~,
\label{eq:markov}
\end{equation}
where
\begin{equation}
R\left(p_i,\{ \sigma_j(t)\}\right) = -\mu p_i(t) +\sum_{n=1}^{k}\sum_{\{\sigma_j(t)\}} {\Lambda}_n 
P\left(0_i(t),\left\{\sigma_j(t)\right\}\right)\delta_{n_i(t),n}~,
\label{eq:rate-general}
\end{equation}
is the rate function  with ${\Lambda}_n $  obeying Eq.~\eref{eq:Lambda-definition}.
The first contribution to $R$ is the probability for node $i$  to recover per time step $\delta t$, i.e. the recovery rate.
The second contribution is the infection probability per unit time and it is proportional to 
the probability $P\left(0_i(t),\left\{\sigma_j(t)\right\}\right) \equiv P\left(\sigma_i(t)=0,\left\{\sigma_j(t)\right\}_{j=1}^k\right)$ for node $i$ to be  in state $S$ and its neighbours in a configuration with states $\left\{\sigma_j(t)\right\}_{j=1}^k$  at time $t$. The total rate of infection is accounted for by summation over all the possible configurations of the neighbourhood, $\left\{\sigma_j(t)\right\}$, for all the possible values of the number of infected neighbours, $n$ (with $\delta_{n_i(t),n}$ being the Kronecker-delta).

  For a particular example of the synergistic SIS process on a $3$-regular graph, the double-summation term multiplied by $\delta t$ in Eq.~\eref{eq:rate-general} represents the  probability of infection of susceptible node $i\equiv 0$ by its neighbours ($j=1,2,3$ ) at time $t+\delta t$, i.e.
  $P(T_0\cap 0_0)=\sum_{\left\{C \right\}}P(T_0|C)P(C)$ where the summation is taken over configurations with fixed state of the central node, $\sigma_0(t)=0$, and all possible $\sigma_j(t)$ for its neighbours, i.e. explicitly,
\begin{eqnarray}
&\delta t \sum_{n=1}^{3}\sum_{\{\sigma_j(t)\}} {\Lambda}_n 
  P\left(0_i(t),\left\{\sigma_j(t)\right\}\right)\delta_{n_i(t),n}=  \sum^3_{n=0}P(T_0|C_n)P(C_n)
\nonumber \\
  &=
  \sum^3_{n=0}(1-(1-\lambda_n \delta t)^n)P(C_n) 
~,
\label{eq:rate-general-2}
\end{eqnarray}
where $C_n$ is a configuration with $n$ infected neighbours of non-infected node $0$, i.e.  $C_0=0_{0}\cap 0_{1}\cap 0_{2}\cap 0_{3}$ for no infected neighbours, $C_1=0_0\cap((1_{1}\cap 0_{2}\cap 0_{3})\cup (0_{1}\cap 1_{2}\cap 0_{3})\cup (0_{1}\cap 0_{2}\cap 1_{3}))$  
for one infected neighbour,
$C_2=0_0\cap((1_{1}\cap 1_{2}\cap 0_{3})\cup (1_{1}\cap 0_{2}\cap 1_{3})\cup (0_{1}\cap 1_{2}\cap 1_{3}))$ for two infected neighbours, and
$C_3=0_0\cap 1_{1}\cap 1_{2}\cap 1_{3}$ for three infected neighbours. 
The probabilities of configurations with particular number of infected nodes  are given by standard expressions, e.g.
$P(C_2)=P(0_0\cap 1_{1}\cap 1_{2}\cap 0_{3})+P(0_0\cap 1_{1}\cap 0_{2}\cap 1_{3}) +P(0_0\cap 0_{1}\cap 1_{2}\cap 1_{3}) $.

  Eq.~(\ref{eq:rate-general}) for marginal probabilities $p_i$ can be obtained from the full set of $2^N$ rate equations for the state probabilities $P\left(\left\{\sigma_i(t) \right\}_{i=1}^N\right)$, which are similar to Eq.~\eref{eq:markov},  by summing up all the equations with fixed value of $\sigma_i(t)=1$ on the left-hand side of these equations. 
 The sum taken over all state variables equals zero, reflecting the conservation of probability while 
 the partial sum with fixed $\sigma_i(t)=1$ leads to a lot of cancellations for the rate terms (on the right-hand side of these equations),  describing transition events occurring away from the neighbourhood of node $i$ so that the  surviving terms describe only the events in which the nearest neighbours of $i$ are involved.
 The rate equations similar to Eq.~(\ref{eq:rate-general}) but without synergy effects are used for description of spreading processes, especially in dealing with dynamical correlations~
 \cite{Silva_Oliveira_2011:JPhysA,Ferreira_2013:EurPhysJ,Pianegonda_Fiore_2015:JStatMech}. 

The model introduced above is well defined for any $\delta t < 1/\lambda_n$ but we will focus on two dynamics:
\begin{itemize}
\item[(i)] Discrete-time dynamics (\emph{d-time}): The state of nodes changes in discrete time steps, $\delta t$, and they do so simultaneously (i.e. updates are synchronous). We set $\delta t=1$ for d-time dynamics in the sequel which implies that the rates coincide with probabilities.
  \item[(ii)] Continuous time dynamics (\emph{c-time}): The time step is infinitesimal, i.e. $\delta t \rightarrow d t$. Accordingly, $\delta p_i \rightarrow d p_i$ and  ${\Lambda}_n=n \lambda_n$. 
\end{itemize}

\subsection{Forms of synergy}

In our model, the synergistic effects are incorporated in the individual transmission rates, $\lambda_n$.
Two particular synergistic mechanisms are analysed below: S-synergy and I-synergy.

For S-synergy, the $k-n_i$ susceptible neighbours of node $i$ multiplicatively affect the individual transmission rates. In general, this can be represented by an exponential functional form, i.e.
\begin{eqnarray}
\lambda_{n_i} = \min \{\alpha e^{\beta(k-n_i)},1/\delta t\}~,  
\label{eq:lambda_n_Ssynergy}
\end{eqnarray}
with $\alpha$ being  the inherent transmission rate  when all the neighbours of $i$ are infected ($n_i=k$) and there are no susceptible neighbours of $i$ which could affect the strength of the attack by the infected neighbours.
The parameter $\beta$ controls the synergy strength of susceptible neighbours.
Positive values of $\beta$ correspond to constructive synergy in which the susceptible neighbours of $i$ encourage the transmission.
In contrast,  negative values represent situations in which susceptible neighbours of $i$ multiplicatively cooperate to prevent transmission to $i$.  
For example, the transmission towards a node $i$ of degree $k>1$ is reduced by a factor $e^\beta$ when there is only one susceptible node connected to $i$. Similarly, if two neighbours of node $i$ are in susceptible state, the greater support leads  to the reduction of infection rate by factor $\left(e^\beta\right)^2$, etc.
The minimum condition in Eq.~\eref{eq:lambda_n_Ssynergy} ensures that the probability of transmission in time step $\delta t$ is at most 1 for any $\alpha$ and $\beta$. 
This form of synergy was proposed as an important factor for the spread of social content~\cite{Gomez-Gardenes_Lotero_Taraskin_FJPR2015}.
Other functional forms of $\lambda_n$ can be used as well although the main features of the synergistic SIS processes are expected to be qualitatively similar (see e.g.~\cite{Perez_Reche_2011:PRL,Taraskin-PerezReche_PRE2013_Synergy,Gomez-Gardenes_Lotero_Taraskin_FJPR2015} where a linear dependence of $\lambda_n$ on $n$ corresponding to additive cooperation was studied). 

For the second type of synergy, I-synergy, the individual  transmission rates are multiplicatively affected by $n_i$ infected  neighbours of susceptible node $i$  which the infection is attempted to be passed to~\cite{Perez_Reche_2011:PRL}, i.e. 
\begin{eqnarray}
\lambda_{n_i} = \min\{\alpha e^{\beta(n_i-1)},1/\delta t\}~. 
\label{eq:lambda_n_Isynergy}
\end{eqnarray}
Here, $\alpha$ is the inherent rate of infection corresponding to the case in which transmission to $i$ comes from a single infected neighbour, i.e. when  $n_i=1$. The synergy parameter $\beta$ accounts for the strength of cooperation ($\beta>0$) or interference ($\beta<0$) between the infected neighbours of $i$. 

  It follows from comparison of Eqs.~\eref{eq:lambda_n_Ssynergy} and \eref{eq:lambda_n_Isynergy} that 
  I-synergy is similar to  S-synergy with inverse sign of the synergy strength, i.e. with $\beta$ replaced by $-\beta$ in Eq.~\eref{eq:lambda_n_Ssynergy} although with a significant distinction.
  Indeed, Eq.~\eref{eq:lambda_n_Ssynergy} can be rewritten as $\lambda_{n_i}= \left(\alpha e^{-\beta(1-k)} \right)e^{-\beta(n_i-1)}$ so that it has the form of Eq.~\eref{eq:lambda_n_Isynergy} but with inherent transmission rate $\left(\alpha e^{-\beta(1-k)} \right)$ dependent on the node degree $k$.
  In contrast, the inherent transmission rate for I-synergy in Eq.~\eref{eq:lambda_n_Isynergy}  is the same for all nodes independent of their degree.
  This difference between two types of synergy can be especially significant in heterogeneous networks where the nodes have different degrees. 

  For both types of synergy, the individual transmission rates are continuous functions of the discrete variable $n_i$ giving the number of infected neighbours. This is in contrast to the threshold models where the transmission rates are described by discontinuous (threshold) functions of $n_i$~\cite{Liu_2007:PRL,Silva_Oliveira_2011:JPhysA,Varghese_2013:PRE,Chae_2015:New_J_Phys,Bottcher_2017:Sci_Rep,Bottcher_2017:PRL,Min_2018:Sci_Rep}.
  The synergistic individual transmission rates  are determined by two parameters, i.e. by the inherent transmission rate $\alpha$ and by the  synergy strength $\beta$, and their dependence on $\alpha$ and $\beta$ is described by continuous functions. 
  In the synergy-free case, i.e. for $\beta=0$, these rates do not depend on $n_i$, and just coincide with the inherent transmission rate.
 All these properties of the synergistic transmission rates make possible to investigate the influence of cooperative effects on SIS process in the whole $(\beta,\alpha)$ parameter space and  reveal, as shown below, quite a rich behaviour even for the simplest case of the SIS spread on k-regular graphs. 

\section{Methods}
To analyse the proposed model, one can either simulate the process numerically by means of Monte Carlo (MC) sampling of the trajectories of the system or through analytical approaches after making certain approximations~\cite{Pastor-Satorras_15:review}. Both procedures were employed to analyse the model. Before presenting the results of our analysis, we describe the ingredients of our analytical calculations.

The  analytical results are based on two standard approximations~\cite{Gleeson_2013:PRX,Pastor-Satorras_15:review}: the single-site approximation and mean-field approximation (sometimes they both are called just a mean-field approximation~\cite{Gleeson_2012:PRE}).
The single-site approximation neglects  dynamical correlations, i.e. 
$
P\left(0_i,\left\{\sigma_j(t)\right\}\right)=(1-p_i)\prod_{j=1}^{k_i}\pi(\sigma_j(t))~,
$ 
where $\pi(\sigma_j(t)=1)=p_j$ and $\pi(\sigma_j(t)=0)=1-p_j$. 

The homogeneous mean-field approximation assumes that the probability of a node being infected does not depend on the node, i.e. $p_i=p$ for all $i$. Despite the fact that this approximation neglects fluctuations completely, it is reasonable 
because, in the thermodynamic limit, random k-regular graphs do not contain loops and in finite networks the number of such loops is exponentially small~\cite{Bollobas2001}.
Combining these two approximations, the rate function (see Eq.~\eref{eq:rate-general}) becomes:
\begin{equation}
R(p_i,\{ \sigma_j(t)\})\simeq R(p) \equiv -\mu p+(1-p) q(p)~,
\label{eq:markov_mean_1}
\end{equation}
where
\begin{equation}
  q(p)=\sum_{n=1}^{k} {k \choose n} {\Lambda}_n p^n (1-p)^{k-n}
  \label{eq:q_mf}
\end{equation}
is the rate at which a susceptible individual gets infected by its infected neighbours.  

A closed form can be found for $q(p)$ for both I- and S-synergy in the c-time limit:
\begin{equation}
  q(p)=\cases{
     \alpha k p \left[(1-p)e^{\beta}+p \right]^{k-1}~, & \rm{for S-synergy}\\
\alpha k p\left[1-p+pe^{\beta} \right]^{k-1}~, & \rm{for I-synergy}}
  \label{eq:q_mf_c-time}
\end{equation}

Note that the heuristic expression $q(p)=\alpha k p e^{\beta k (1-p)}$ proposed in \cite{Gomez-Gardenes_Lotero_Taraskin_FJPR2015} for d-time S-synergy differs from the expressions derived here  more rigorously for both c- and d-time dynamics. The expression given in \cite{Gomez-Gardenes_Lotero_Taraskin_FJPR2015} coincides with those given here for non-synergistic spread ($\beta =0$) and captures the key trends when $\beta \ne 0$. For instance, it is a monotonically increasing function of $p$ for any $\alpha, k >0$. Strictly speaking, however, one should use Eq.~\eref{eq:q_mf} in order to describe accurately  the synergistic effects of any transmission rate within the single-site mean-field approximation.

\section{Results}

The behaviour of the process in the long-time limit $t\to \infty$ is of special interest because it determines if the network is vulnerable to the infection spread or not. We will analyse the effect of synergistic transmission in this limit in which the SIS process reaches a stationary (quasi-stationary in finite systems) regime such that $R(p_i,\{ \sigma_j(t)\})=0$.

Due to the non-linear nature of the rate function, the number of its fixed points can vary when the parameters of the model change and different stationary regimes  can emerge in the parameter space. For c-time dynamics and given node degree, stationary states can be fully parametrised with two parameters, $\tilde{\alpha}=\alpha/\mu$  and $\beta$ since $R(p)$ is linear in $\lambda_n$ (this is clear from Eq.~\eref{eq:rate-general} and expression  ${\Lambda}_n=n \lambda_n$ for c-time). In contrast, the stationary states for d-time processes depend on $\alpha$, $\beta$ and $\mu$. This is an important difference compared to the stationary states for non-synergistic SIS processes that depend on $\tilde{\alpha}$ for c-time  and $\tilde{\alpha}$ and $\mu$ for d-time  dynamics.  

Different stationary regimes are separated in the parameter space by bifurcations at which the number of stationary points (roots of $R(p)$)  changes~\cite{Strogatz_NonlinearBook}. The results presented below involve bifurcations of three different types:
\begin{itemize}
	\item[(i)] ~Transcritical (TC) bifurcation.
	This is a codimension-one bifurcation point and it corresponds to double root at $p_0=0$ with condition, $R(0)=R'(0)=0$. 	The infection probability $p$ changes continuously when a TC bifurcation is crossed in the parameter space.
	\item[(ii)] ~Saddle-node (SN) bifurcation.
	This is a codimension-one bifurcation point and it corresponds to double root at finite density, $p_*$, with conditions, $R(p_*)=R'(p_*)=0$, for $p_*\in(0,1)$. The stationary density changes discontinuously when a SN point is crossed.
	\item[(iii)] ~~Saddle-node-transcritical (SNT) crossing bifurcation. 
	This is a codimension-two bifurcation point and it corresponds to triple root at zero density with conditions, $R(0)=R'(0)=R''(0)=0$. 	The stationary density near zero changes continuously  when an SNT point is crossed. 
\end{itemize}

\subsection{S-synergy}

\begin{figure}
  \centering
  {\includegraphics[clip=true,width=0.45\textwidth]{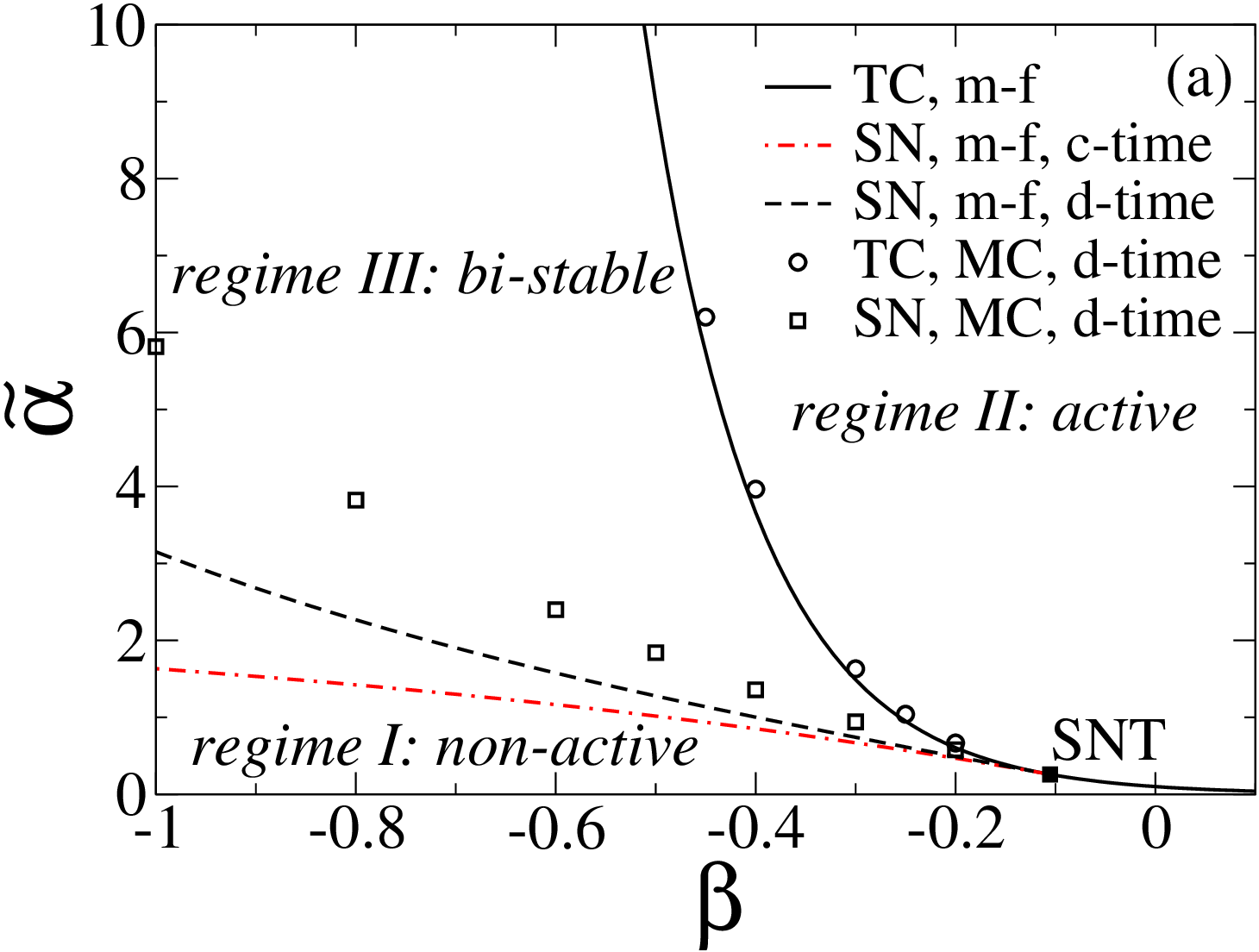}}
  \quad
  {\includegraphics[clip=true,width=0.45\textwidth]{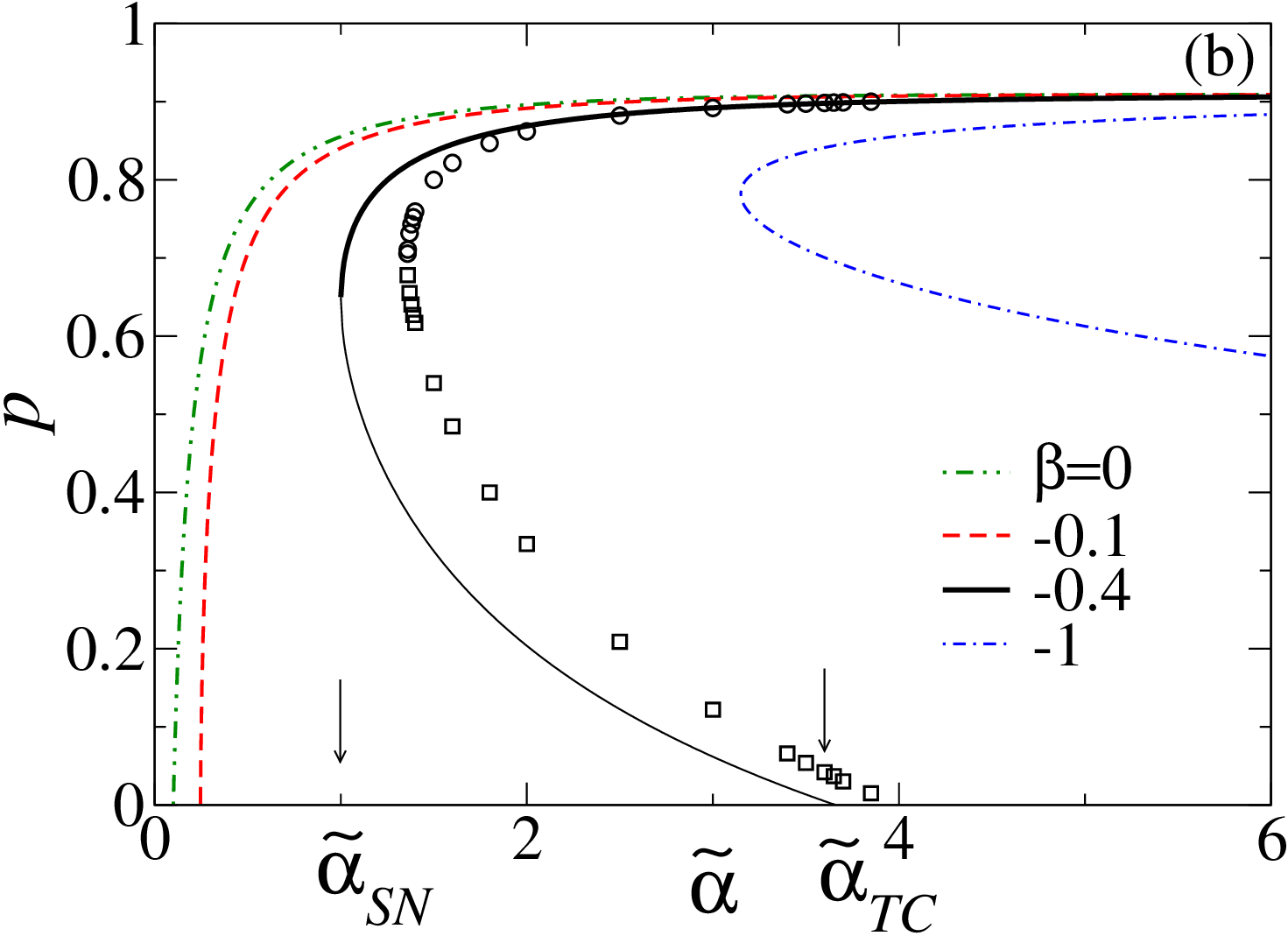}}
  \quad
  {\includegraphics[clip=true,width=0.45\textwidth]{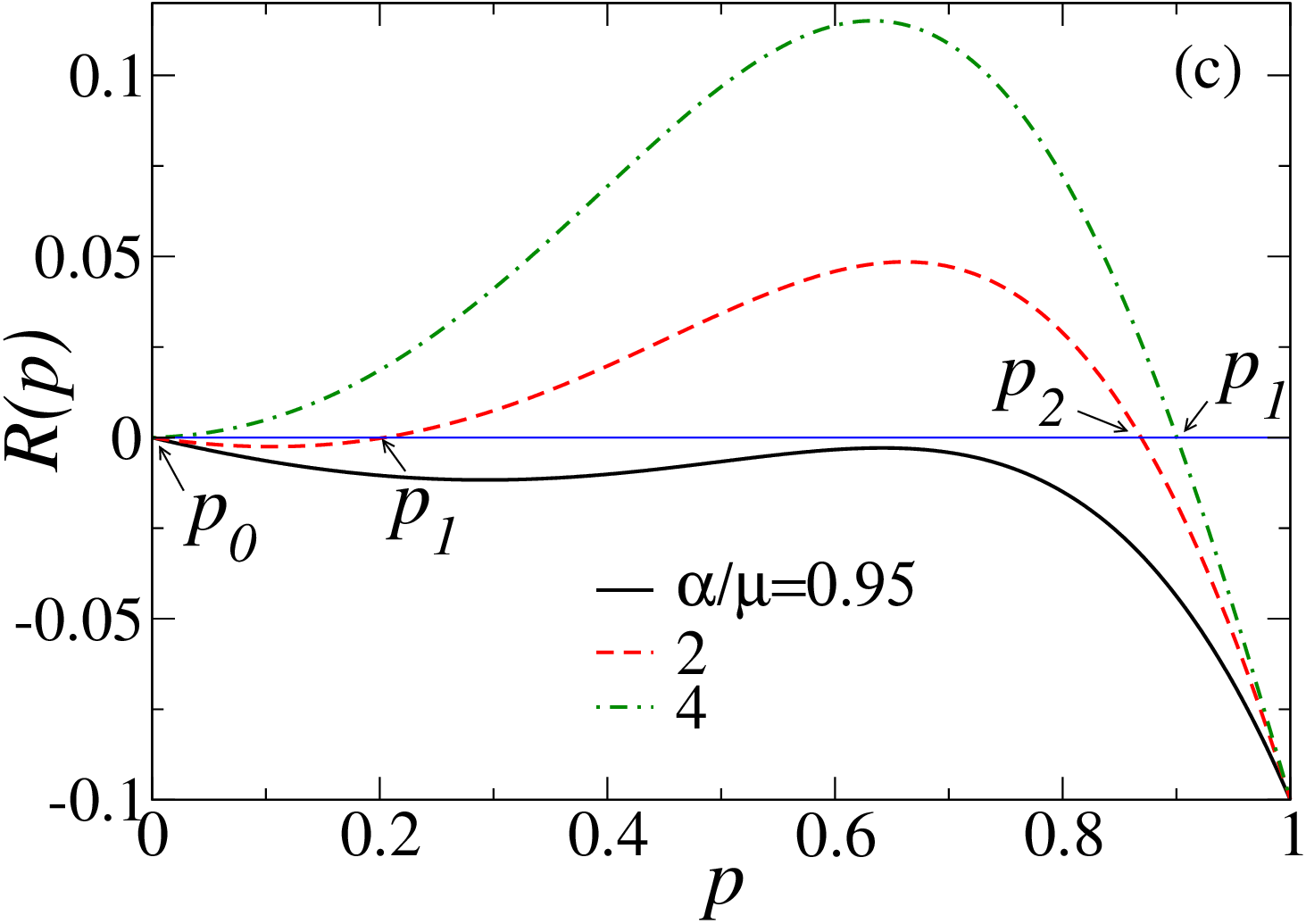}}
   \quad
  {\includegraphics[clip=true,width=0.45\textwidth]{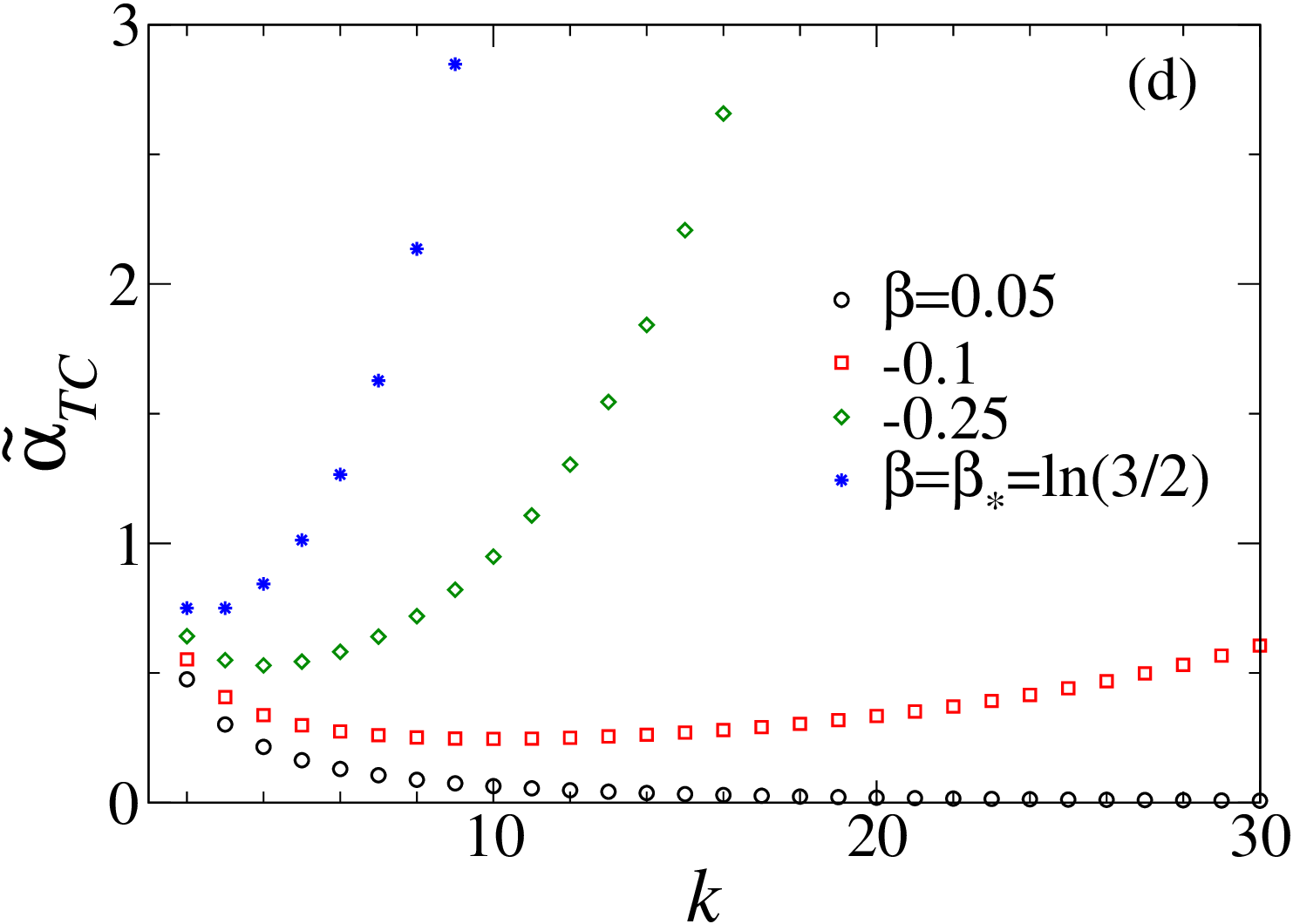}}
	\caption{\label{fig:phase_diag_KR_10_S} 
          {\bf (a)}~Bifurcation diagram for SIS processes with S-synergy on a random k-regular graph with $k=10$ following c- and d-time dynamics.    The solid line represents the line of TC biburcations given by mean-field (m-f) Eq.~\eref{eq:alpha_1_KR}. The dashed and dot-dashed  lines show SN bifurcations for d-time dynamics with $\mu=0.1$  and c-time dynamics (see Eq.~\eref{eq:alpha_2}),  respectively.
          The circles (TC biburcations) and squares (SN bifurcations) represent the data obtained numerically for random 10-regular graphs of size $N=10^5$ for SIS processes following the rules of d-time dynamics with $\mu=0.1$.
The error bars for MC data are lees than the symbols size.          The solid square shows the SNT crossing bifurcation  with coordinates  given by Eqs.~\eref{eq:snt_alpha}-\eref{eq:snt_beta} for c-time dynamics.
         {\bf  (b)}~Dependence of the concentration of infected nodes $p$ in quasi-equilibrium state for SIS process with S-synergy on random 10-regular graph ($N=10^5$)   {\it vs}   relative transmission probability $\tilde{\alpha}$ for d-time dynamics with   $\mu=0.1$ and different values of the synergy parameter $\beta$ shown in the legend. In the bi-stable regime for $\beta=-0.4$, the thick and thin solid lines correspond to finite-density stable and unstable stationary states of the synergistic SIS process, respectively.
          Squares (unstable equilibrium) and circles (stable equilibrium) show numerical results for d-time dynamics with $\mu = 0.1$ and $\beta = -0.4$.
         {\bf  (c)}~Dependence of the rate function, $R(p)$, calculated within the single-site mean-field approximation on the concentration of infected nodes, $p$, for d-time SIS process with $\mu=0.1$, $\beta=-0.4$ and different values of  $\tilde{\alpha}=\alpha/\mu$ as indicated in the legend.
          {\bf (d)}~Dependence of the TC bifurcation point, $\tilde{\alpha}_{\rm{TC}}(\beta;k)$ given by Eq.~\eref{eq:alpha_1_KR}, on the node degree $k$ for random k-regular graphs.  
	}
\end{figure}

The location of the bifurcation points in $(\beta,\tilde{\alpha}) $  parameter space, i.e. bifurcation diagram, for SIS process exhibiting S-synergy in transmission on random k-regular graph is shown in Fig.~\ref{fig:phase_diag_KR_10_S}(a) for a typical set of parameters. 
The data presented by lines were obtained withing the single-site mean-field approximation while the open symbols correspond to results of MC simulations (see Sec.~\ref{sec:numerics} for more detail). 
There are three different regimes: regime I (non-active), regime II (active) and regime III (bi-stable).  A general analysis of stability of all fixed points found in the paper is given in~\ref{app:stability}.
In regime I, the synergistic SIS process  is characterised by a single fixed point located at $p_0=0$ which is globally asymptotically stable in the feasible interval with $p\in (0,1]$ (see Corollary~\ref{co:RegimeI}).  The solid line in Fig.~\ref{fig:phase_diag_KR_10_S}(c) shows an example of the rate function $R(p)$ which only has a root at $p=p_0=0$.
In regime II, there are two fixed points located at $p_0 =0$ and $p_1>0$ (see the dot-dashed $R(p)$ curve in Fig.~\ref{fig:phase_diag_KR_10_S}(c)).
The fixed point at $p_0$ is unstable and processes in this regime evolve towards the active regime with  fixed point at $p_1$  which is globally asymptotically stable in the feasible interval (see Corollary~\ref{co:RegimeII}).
Regime III is characterised by three fixed points, two locally stable at $p_0=0$ and $p_2>0$ and one unstable at $p_1 \in (p_0,p_2)$ (see the  dashed curve in Fig.~\ref{fig:phase_diag_KR_10_S}(c) and Corollary~\ref{co:RegimeIII}).
Note that the bifurcation (phase) diagram of the model is qualitatively similar to that presented in~\cite{Gomez-Gardenes_Lotero_Taraskin_FJPR2015} but there are quantitative differences due to  different treatment of the function $q(p)$.

\begin{figure}
  \centering
  {\includegraphics[clip=true,width=0.45\textwidth]{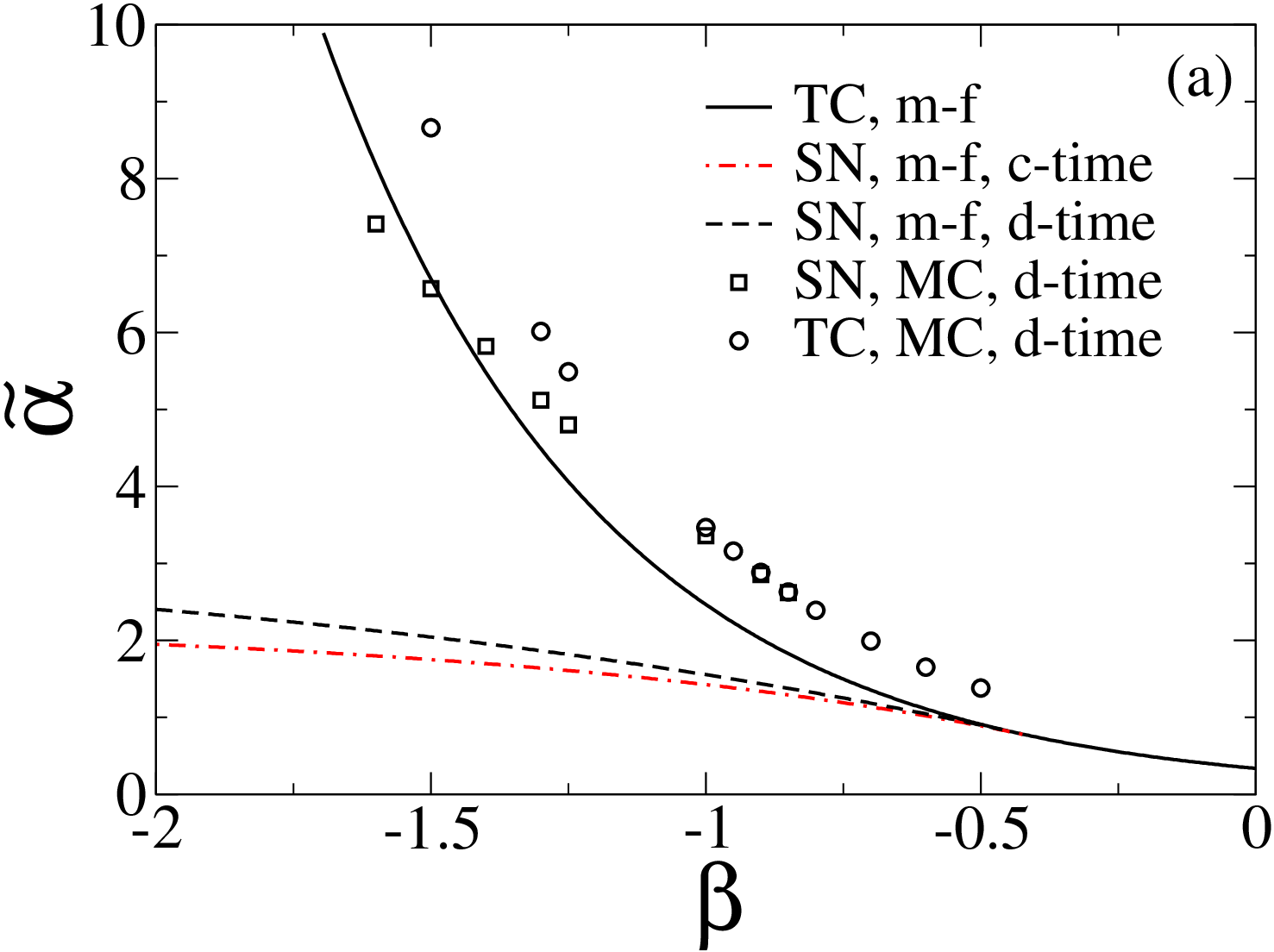}}
  \quad
  {\includegraphics[clip=true,width=0.45\textwidth]{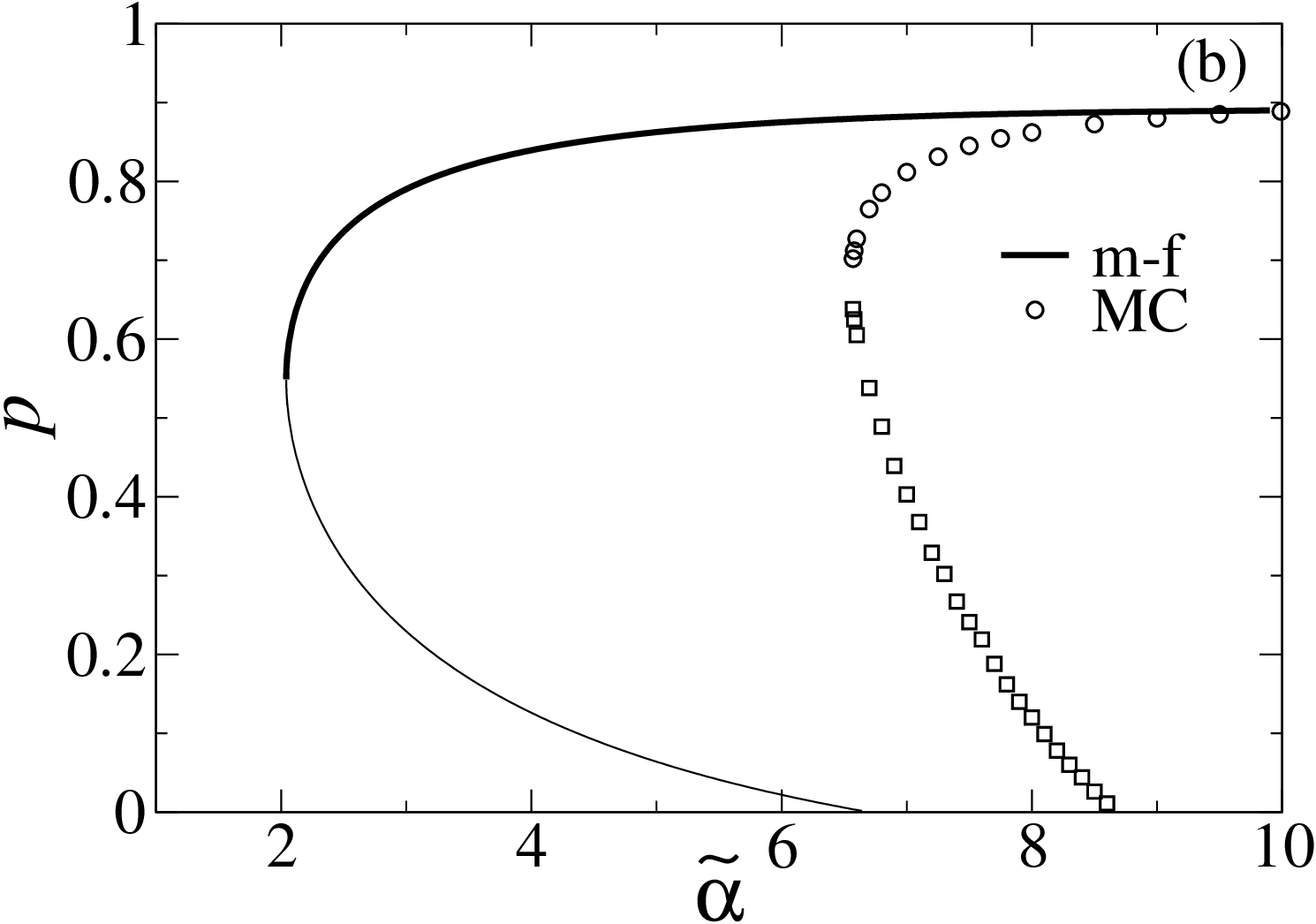}}
	\caption{\label{fig:phase_diag_KR_3_S} 
          {\bf (a)}~Bifurcation diagram for SIS processes with S-synergy on a random $3$-regular graph. The same values of parameters and line and symbol styles as in Fig.~\ref{fig:phase_diag_KR_10_S}(a) are used.  
         {\bf  (b)}~Dependence of the concentration of infected nodes $p$ in quasi-equilibrium state for an SIS process with S-synergy on a random 3-regular graph   {\it vs}   relative transmission probability $\tilde{\alpha}$ for d-time dynamics with $\beta=-1.5$. The mean-field (m-f) and MC simulations data are shown by lines and symbols, respectively. The values of other parameters used in the simulations are the same as those for data shown in Fig.~\ref{fig:phase_diag_KR_10_S}(b). 
	}
\end{figure}

The different regimes are separated by lines of bifurcations. The solid line in Fig.~\ref{fig:phase_diag_KR_10_S}(a), $\tilde{\alpha}=\tilde{\alpha}_{\rm{TC}}(\beta)$,  represents the TC bifurcations separating active and non-active regimes on the right ($\beta>\beta_{\rm{SNT}}$) to the SNT crossing bifurcation (solid square) at  $\left(\beta_{\rm{SNT}},\tilde{\alpha}_{\rm{SNT}}\right)$, and active and bi-stable regimes on the left ($\beta<\beta_{\rm{SNT}}$) to the SNT point, respectively. 
The broken style lines with $\tilde{\alpha}=\tilde{\alpha}_{\rm{SN}}(\beta)$ correspond to the SN bifurcations between non-active and bi-stable regimes.

For relatively large fixed values of $\beta > \beta_{\rm{SNT}}$, the SIS process is in non-active  regime I  for $\tilde{\alpha} < \tilde{\alpha}_{\rm{TC}}$ and in active regime II for $\tilde{\alpha} > \tilde{\alpha}_{\rm{TC}}$. In this case, the probability $p$ for a node to be infected or equivalently  the  concentration of infected nodes  is a single-valued function of $\tilde{\alpha}$ and $\beta$ (see the dashed and dot-dashed curves in Fig.~\ref{fig:phase_diag_KR_10_S}(b)).
Within the single-site mean-field approximation, the functional form for the TC bifurcation line and coordinates of the SNT crossing bifurcation can be found analytically both for d- and c-time dynamics. 
Indeed, the TC bifurcation line where $R'(0)=0$ satisfies 
\begin{eqnarray}
\tilde{\alpha}_{\rm{TC}}(\beta)=\frac{1}{k} e^{-\beta(k-1)}~,
\label{eq:alpha_1_KR}
\end{eqnarray}
for both d- and c-time dynamics. This line is independent of $\mu$ and reduces to the well-known threshold condition $\tilde{\alpha}_{\rm{TC}}(\beta)=1/k$ for non-synergistic epidemics with $\beta=0$~\cite{Pastor-Satorras_15:review,Barrat_08:book}. 

The coordinates of the SNT crossing bifurcation point (solid square in Fig.~\ref{fig:phase_diag_KR_10_S}(a)), where  $R(0)=R'(0)=R''(0)$, are given by the following expressions,   
\begin{equation}
\tilde{\alpha}_{\rm{SNT}}=\cases{
     \frac{1}{k}\left[\frac{k}{\mu}\left(1-\sqrt{ 1 -\frac{2\mu}{k-1}} \right) \right]^{k-1}~,&\rm{for d-time}\\
\frac{1}{k}\left(\frac{k}{k-1} \right)^{k-1}~,&\rm{for c-time}}
\label{eq:snt_alpha}
\end{equation}
\begin{equation}
  \beta_{\rm{SNT}}=\cases{
     - \ln \left[\frac{k}{\mu}\left(1-\sqrt{ 1 -\frac{2\mu}{k-1}} \right) \right]~,&\rm{for d-time}\\
\ln\left(\frac{k-1}{k} \right)~,&\rm{for c-time.}}
 \label{eq:snt_beta}                                  
\end{equation}
Note that the coordinates of the SNT crossing bifurcation for c-time dynamics   are obtained  by the limit for $\mu\to 0$ in the expressions for d-time dynamics.

For sufficiently negative fixed values of $\beta < \beta_{\rm{SNT}}$ (to the left of the SNT crossing bifurcation point in Fig.~\ref{fig:phase_diag_KR_10_S}(a)), the behaviour of the synergistic SIS processes changes drastically.
In particular, $p$  becomes a multivalued function of $\tilde{\alpha}$ in the interval $(\tilde{\alpha}_{\rm{SNT}}(\beta), \tilde{\alpha}_{\rm{TC}})$ where the SIS process is in bi-stable regime III (see the multi-valued function $p(\tilde{\alpha})$ shown by the solid line  in Fig.~\ref{fig:phase_diag_KR_10_S}(b)). In the bi-stable regime,  the equilibrium concentration of infected nodes depends on initial concentration of infected nodes, $p(0)$. 
Within mean-field, the SIS process is non-active if $p(0)<p_1$, where $p_1$ is the middle root of $R(p)$. In contrast, it is active if  $p_1<p(0)$, i.e. it reaches a finite concentration $p=p_2>0$, where $p_2$ is the largest root of $R(p)$.

The expression for $\tilde{\alpha}_{\rm{SN}}(\beta)$ corresponding to  the line of SN bifurcations can  be found analytically only for c-time dynamics (the result for d-time dynamics can be obtained numerically). Indeed, using the condition  $R(p_*)=R'(p_*)=0$  and Eq.~\eref{eq:q_mf_c-time} for c-time S-synergy, one obtains
 \begin{equation}
 \tilde{\alpha}_{\rm{SN}}(\beta)=\left(\frac{k}{k-1}\right)^{k-1} \left(1-e^{\beta}\right)~
\label{eq:alpha_2}
\end{equation}
 which is valid for $\beta \le \beta_{\rm{SNT}}$. 
 For d-time dynamics, the shape of the SN bifurcation line  depends on the value of $\mu$ and tends to the SN bifurcation line for c-time dynamics (the dot-dashed line in Fig.~\ref{fig:phase_diag_KR_10_S}(a)) when $\mu\to 0$. 

 We analysed above the behaviour of SIS processes in a representative case of random k-regular graphs with $k=10$.
 However, all the qualitative findings hold for other values of $k$ as well.
 As an example, in Fig.~\ref{fig:phase_diag_KR_3_S}, we show the bifurcation diagram and dependence of concentration of infected nodes on the reduced transmission rate, $\tilde{\alpha}$, for 3-regular graph.
   As seen from comparison of Fig.~\ref{fig:phase_diag_KR_3_S} with Fig.~\ref{fig:phase_diag_KR_10_S}, qualitatively the behaviour of the synergistic SIS process is the same in both 3- and 10-regular graphs. 
   However, the quality of the single-site mean-field approximation in description of the synergistic SIS processes studied numerically becomes noticeably  better  with increasing $k$.
   This is a known effect according to which the states of two neighbouring nodes are 
      less likely to depend on the state of each other if they are connected to many other neighbours which are more likely to influence the pair of neighbours~\cite{Gleeson_2012:PRE,Gleeson_2013:PRX}. 

   The effect of the node degree on different regimes exhibited by the model can be readily analysed using the analytical expressions for bifurcations. We restrict our analysis to $k\geq 2$ to ensure a giant connected component in the network. 
The region corresponding to regime III in the $(\beta,\tilde{\alpha})$ space increases with $k$, i.e. bi-stable behaviour is overall more likely when increasing the connectivity of the network. Indeed, for $\beta<\beta_{\rm{SNT}}$, the relative inherent transmission rate $\tilde{\alpha}_{\rm{SN}}(\beta;k)$ exhibits a mild increase with $k$ that is counteracted by a faster increase of $\tilde{\alpha}_{\rm{TC}}(\beta;k)$. This results in an enlargement of the region for bi-stable behaviour with increasing $k$.

The dependence of $\tilde{\alpha}_{\rm{TC}}(\beta; k)$ on $k$ is more interesting. Depending on the value of $\beta$, one can distinguish three cases (see Fig.~\ref{fig:phase_diag_KR_10_S}(d)). 

\begin{itemize}
\item[(i)] $\beta < \beta_* \equiv \ln (2/3)$: One can prove (bearing in mind that $\tilde{\alpha}_{\rm{TC}}(\beta_*;k=2)=\tilde{\alpha}_{\rm{TC}}(\beta_*;k=3)$; see the stars in Fig.~\ref{fig:phase_diag_KR_10_S}(d)) that $\tilde{\alpha}_{\rm{TC}}(\beta;k)$ increases monotonically with $k \geq 2$ . This behaviour can be intuitively understood in terms of a higher opposition of susceptible neighbours to transmission for large $k$. 
  \item[(ii)] $\beta_* < \beta <0$: This is a more counter-intuitive regime since $\tilde{\alpha}_{\rm{TC}}(\beta;k)$ has a minimum at $k=\lceil e^{\beta}(1-e^{\beta})^{-1} \rceil$, where $\lceil \cdot \rceil$ is the ceiling function  (see the squares and diamonds in Fig.~\ref{fig:phase_diag_KR_10_S}(d)). This means  that there is a value of $k$ for which the system is particularly vulnerable to leaving the non-active regime.
\item[(iii)] $\beta >0$: The intrinsic rate $\tilde{\alpha}_{\rm{TC}}(\beta)$ decreases monotonically with $\beta$ (see the circles in Fig.~\ref{fig:phase_diag_KR_10_S}(d))  . This behaviour is again intuitively expected since susceptible neighbours encourage transmission from infected nodes and this makes the invasion more likely (i.e. occurs for smaller values of $\alpha$) if connectivity is large. Transitions in this case are always between regimes I and II.
\end{itemize}

\subsection{I-synergy}

\begin{figure}
  \centering
  \includegraphics[clip=true,width=0.48\textwidth]{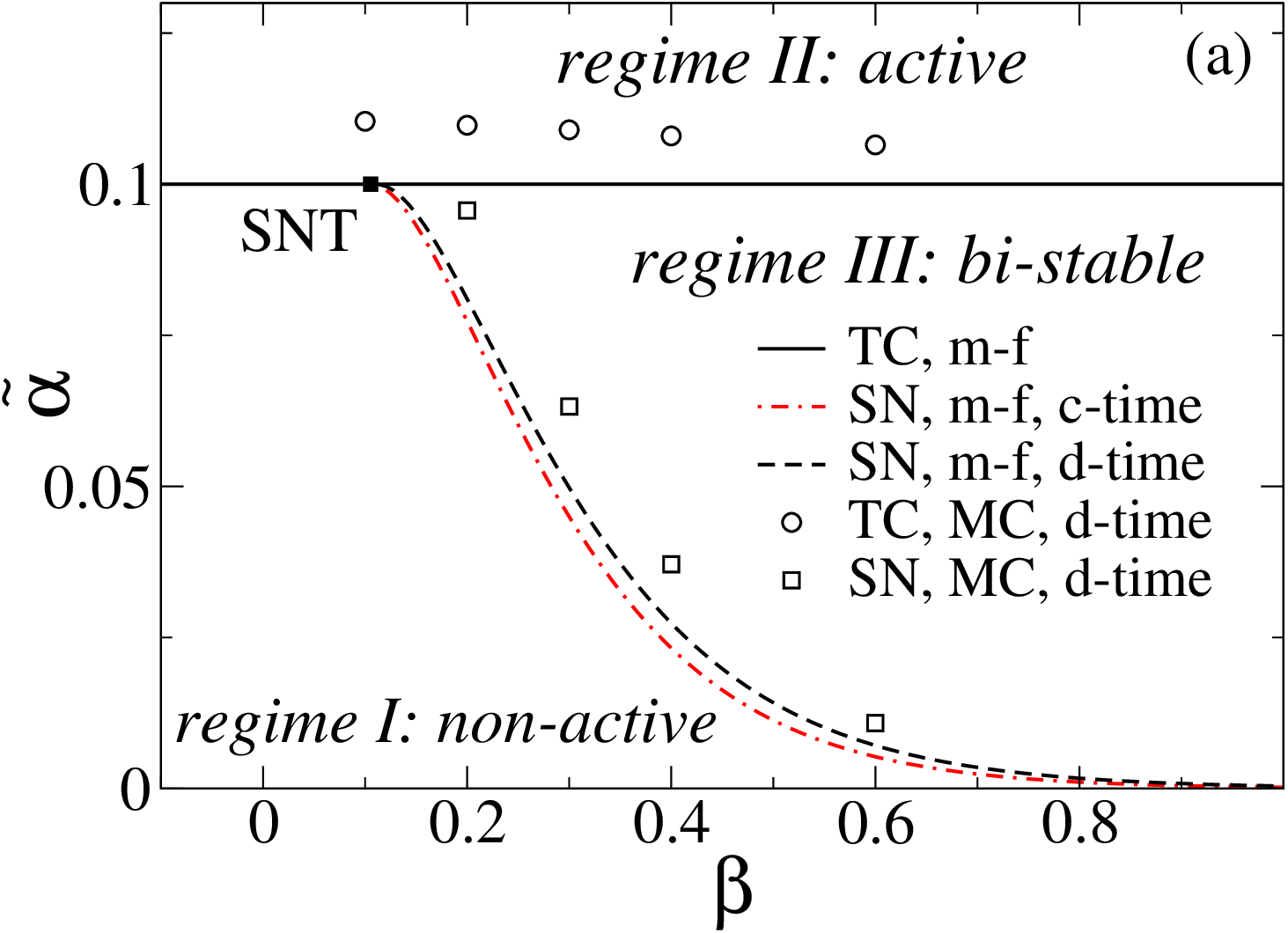}
\quad
\includegraphics[clip=true,width=0.48\textwidth]{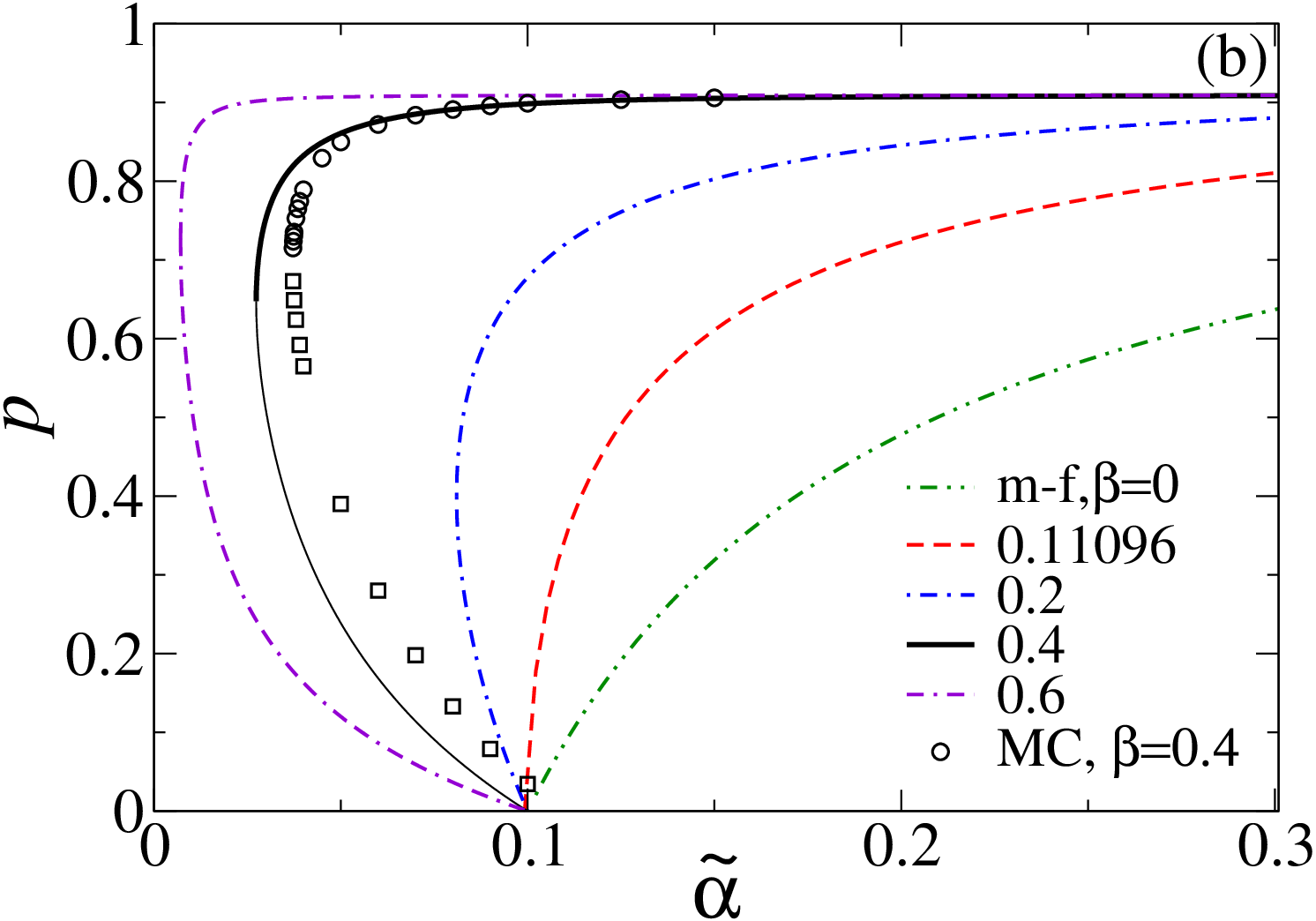}
	\caption{\label{fig:phase_diag_KR_10_I}
 {\bf (a)}~Bifurcation diagram for SIS processes with I-synergy on random k-regular graphs with $k=10$ following c- and d-time dynamics.    The solid line represents the line of TC biburcations given by Eq.~\eref{eq:alpha_1_Isynergy}. The dashed and dot-dashed  lines show SN bifurcations for d-time dynamics with $\mu=0.1$  and c-time dynamics (see Eq.~\eref{eq:alpha_2_Isynergy}),  respectively.
          The circles (TC biburcations) and squares (SN bifurcations) represent the data obtained numerically for random 10-regular graphs of size $N=10^5$ for SIS processes following the rules of d-time dynamics with $\mu=0.1$. The solid square shows the SNT crossing bifurcation  with coordinates  given by Eqs.~\eref{eq:snt_alpha_I}-\eref{eq:snt_beta_I} for c-time dynamics.
          {\bf (b)}~Dependence of the concentration of infected nodes $p$ in quasi-equilibrium state for SIS process with I-synergy on random  10-regular graph ($N=10^5$)   {\it vs}   relative transmission probability $\tilde{\alpha}$ for d-time dynamics with   $\mu=0.1$ and different values of the synergy parameter $\beta$ shown in the legend. In bi-stable regime for $\beta=-0.4$, the thick and thin solid lines correspond to finite-density stable and unstable stationary states of the synergistic SIS process, respectively.  Squares (unstable stationary state)  show numerical results for d-time dynamics $\mu = 0.1$ and $\beta = -0.4$.		
	}
\end{figure}

Fig.~\ref{fig:phase_diag_KR_10_I}(a) shows a typical phase diagram for SIS processes exhibiting I-synergy in transmission and spreading on  random k-regular  graphs with $k=10$. As in the case of S-synergy, there are three different regimes: regime I (non-active), regime II (active) and regime III (bi-stable). As follows from this figure, only constructive I-synergy affects the SIS processes making them more invasive as compared to the synergy-free case. 
Indeed, if $\beta > \beta_{\rm{SNT}}$ then even for relatively small inherent infection rates, i.e. $\tilde{\alpha} < \tilde{\alpha}_{\rm{TC}}$ (but $\tilde{\alpha} > \tilde{\alpha}_{\rm{SN}}$) the network becomes vulnerable for invasion and the concentration of infected nodes can increase abruptly from zero to a finite value (see the curves in Fig.~\ref{fig:phase_diag_KR_10_I}(b) for $\beta > \beta_{\rm{SNT}}\simeq 0.11096$).  
The location of bifurcation points can be found within the single-site mean-field approximation following similar steps as above for S-synergy.
A TC bifurcation occurs at an inherent rate given by 
\begin{eqnarray}
\tilde{\alpha}_{\rm{TC}}(\beta)=\frac{1}{k}
~,  
\label{eq:alpha_1_Isynergy}
\end{eqnarray}
which is not affected by synergy (does not depend on $\beta$; see the horizontal line in Fig.~\ref{fig:phase_diag_KR_10_I}(a)) and coincides with that for the synergy-free case~\cite{Pastor-Satorras_15:review,Barrat_08:book}. 
This is because it is determined by transmission events with a single attacker only. 
Despite not affecting $\tilde{\alpha}_{\rm{TC}}(\beta)$, constructive I-synergy can affect the SIS processes so that they become bi-stable (see Fig.~\ref{fig:phase_diag_KR_10_I}(a)) in a similar way to the SIS processes with S-synergy.  
A  single SNT crossing bifurcation is present in the bifurcation diagram and its coordinates can be obtained analytically for both c- and d-time dynamics:
\begin{eqnarray}
  &\tilde{\alpha}_{\rm{SNT}}=\frac{1}{k},\quad\quad {\hskip148pt} \textrm{for\; both\; c- \; and\; d-time}
\label{eq:snt_alpha_I}
  \\
  &\beta_{\rm{SNT}}=\cases{
     \ln \left[\frac{k}{\mu}\left(1-\sqrt{ 1 -\frac{2\mu}{k-1}} \right) \right]~,&\rm{for d-time}\\
\ln\left(\frac{k}{k-1} \right)~,&\rm{for c-time}}
      \label{eq:snt_beta_I}                            
\end{eqnarray}

The expressions for both the value of the elementary rate, $\tilde{\alpha}_{\rm{SN}}(\beta)$, and concentration of infected nodes, $p_{\rm{SN}}(\beta)$, can be obtained analytically for c-time dynamics at the SN bifurcation separating non-active and bi-stable regimes for $\beta\ge \beta_{\rm{SNT}}$:  
\begin{eqnarray}
\tilde{\alpha}_{\rm{SN}}(\beta)&=&\left(\frac{k}{k-1}\right)^{k-1} \left(e^{\beta}-1\right)e^{-\beta k}~,
\label{eq:alpha_2_Isynergy} \\
p_{\rm{SN}}(\beta)&=&1-\frac{1}{k(1-e^{-\beta})}~.
\label{eq:pSN_Isynergy}
\end{eqnarray}
In case of d-time dynamics, these quantities can be calculated numerically. 

\subsection{Invasion threshold}
\label{sec:InvasionThreshold}
The existence of a well-defined threshold separating non-active and active regimes is an appealing idea in mathematical epidemiology which is often expressed in terms of the basic reproduction number $R_0$~\cite{HethcoteSIAM2000,Pastor-Satorras_15:review}. A well-defined threshold exists under quite general conditions~\cite{VandenDriessche2002} and corresponds to a TC bifurcation separating regimes I and II with $R_0=1$ at the threshold. Here, we show that synergistic effects restrict the regime of validity of the concept of a threshold defined by condition $R_0=1$. Similar deviations from the normal threshold criterion were found in previous works studying the effect of non-linear incidence rates on invasions~\cite{Gubbins-Gilligan_2000,Liu1987}.

For the models studied here, one can define a basic reproduction number as
\begin{equation}
\label{eq:R0}
 R_0=\frac{k}{\mu} \lambda_1~.
\end{equation}
This formula can be intuitively interpreted as the average number of nodes that become infected at the initial stages of the epidemic by the transmission of infection with rate $\lambda_1$ from an infected node during its infectious period which is of the order of $1/\mu$.

By using $\lambda_1 = \alpha e^{\beta(k-1)}$ for S-synergy and condition (\ref{eq:alpha_1_KR}) for the TC bifurcation, one can express the reproductive number as $R_0=\alpha/\alpha_{\rm{TC}}$. Therefore, the condition for the TC bifurcation along the line $\alpha_{\rm{TC}}(\beta)$ is clearly equivalent to the common threshold condition $R_0=1$. The threshold concept defined by  $R_0=1$ is, however, only meaningful for $\beta \geq \beta_{\rm{SNT}}$. Indeed, for $\beta < \beta_{\rm{SNT}}$, the active regime is already stable in regime III where $\alpha \in (\alpha_{\rm{SN}}(\beta),\alpha_{\rm{TC}}(\beta))$. This implies that, depending on the initial conditions, synergistic invasions are possible for $R_0 > \alpha_{\rm{SN}}(\beta)/\alpha_{\rm{TC}}(\beta)$, i.e. they are possible even if $R_0<1$.

The reproduction number in case of I-synergy can be similarly expressed as   $R_0=\alpha/\alpha_{\rm{TC}}$ but with  $\alpha_{\rm{TC}}=\mu/k$ (cf. Eq.~\eref{eq:alpha_1_Isynergy}). Following a similar reasoning as for S-synergy it is easy to demonstrate that the definition of the threshold by condition $R_0=1$ is only meaningfull for $\beta \leq \beta_{\rm{SNT}}$. For $\beta > \beta_{\rm{SNT}}$, invasions are again possible in regime III where $R_0<1$.

\subsection{A minimal model}
In the presence of synergy, the rate function $R(p)$ given by Eq.~\eref{eq:q_mf} is a polynomial of order $k$. However, we have only found regimes with at most three roots for $R(p)$. This suggests the existence of a simplified normal form for the model which can capture all the three dynamical regimes described above. Indeed, a cubic normal form, $R_{\rm{nf}}$,  for the rate function, 
\begin{equation}
\label{eq:Rnf}
R_{\rm{nf}}(p)=ap+bp^2-cp^3~,
\end{equation}
is sufficient to qualitatively capture all the regimes predicted by the full model for any $k$. A similar normal form was proposed in \cite{Saputra2010} to describe the interaction of TC and SN bifurcations in an extended Lotka-Volterra model.

\begin{figure}
  \centering
  {\includegraphics[clip=true,width=0.45\textwidth]{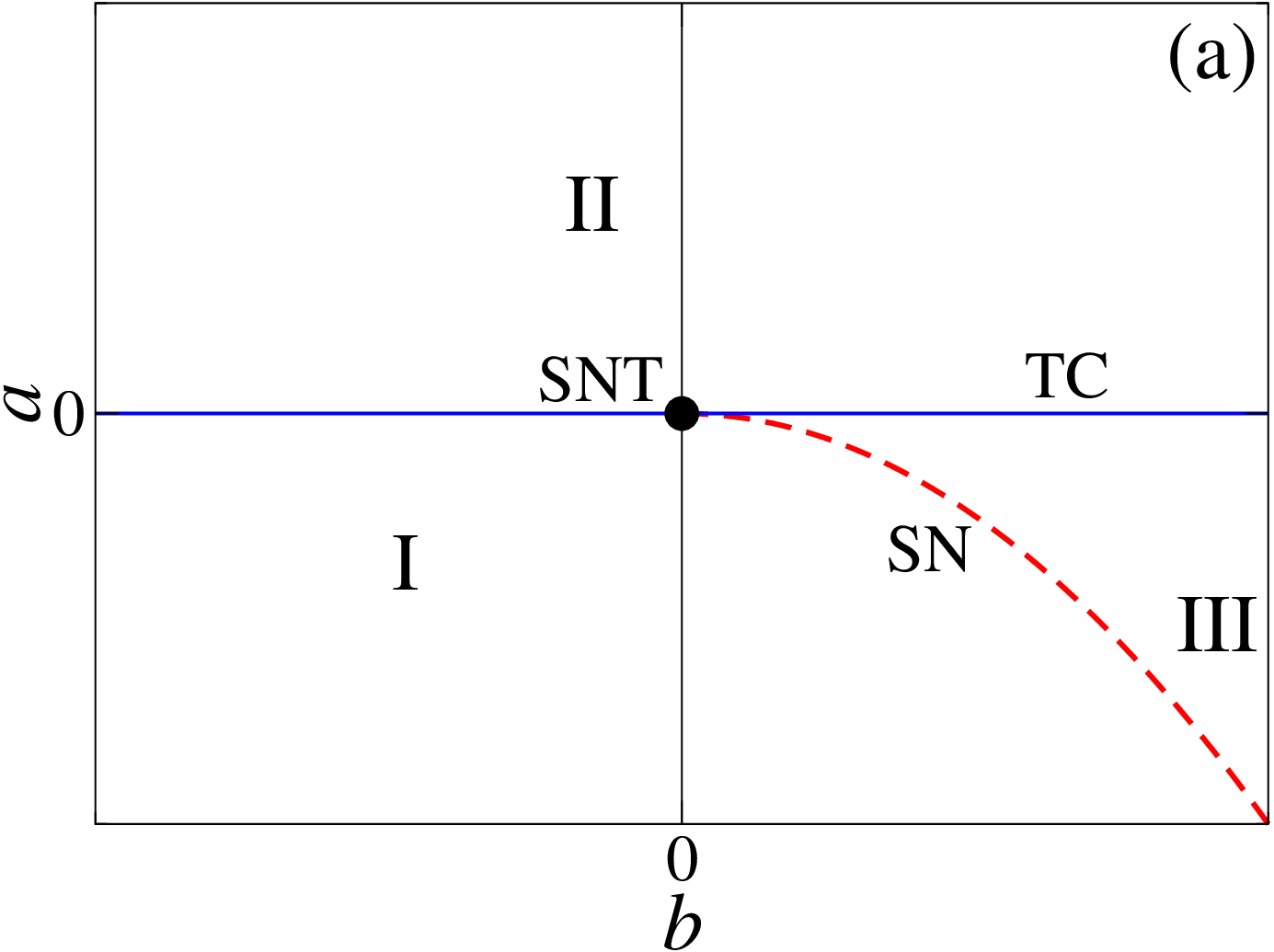}}
  \quad
  {\includegraphics[clip=true,width=0.45\textwidth]{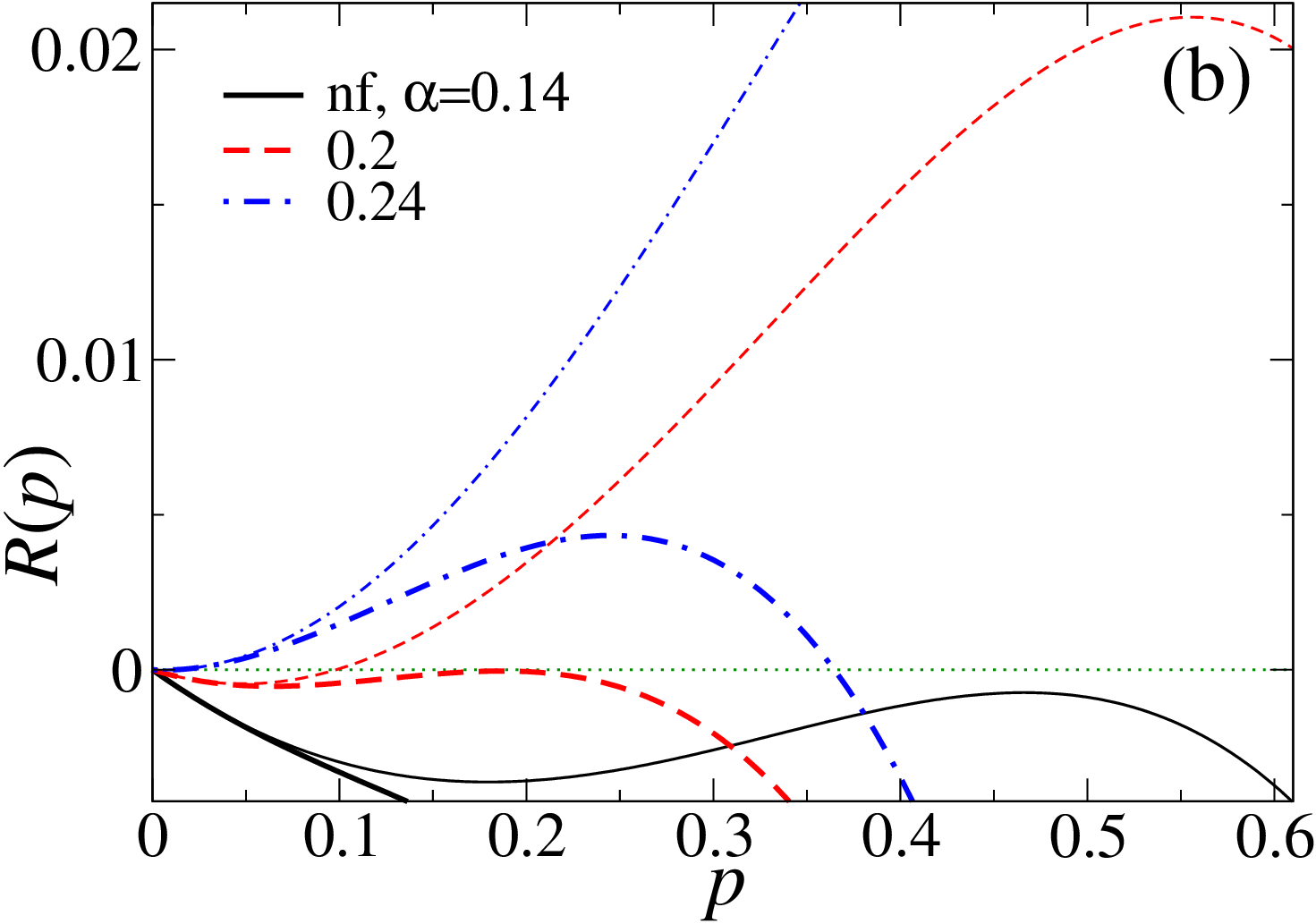}}
  \caption{\label{fig:Normal_Form_NF_R} {\bf (a)}
    Bifurcation diagram for the minimal model based on the normal form for the rate function, $R_{\rm{nf}}(p)$.
   {\bf (b)}   Comparison of the normal form $R_{\rm{nf}}(p)$ (bold lines, Eq.~\eref{eq:Rnf}) and mean-field rate function $R(p)$ (thin lines)  for $k=3$, $\mu = 0.1$, $\beta=-1<\beta_{\rm{SNT}} \simeq -0.43$ and several values of $\alpha$ as indicated in the legend.}
\end{figure}

The normal form is obtained through an expansion of $R(p)$ given by Eq.~\eref{eq:q_mf} up to and including terms $\propto \rm{O}(p^3)$ around $p=0$ which gives the following expressions for  coefficients in Eq.~\eref{eq:Rnf}:
\begin{eqnarray}
\label{eq:a}
  a&=&-\mu+k \Lambda_1~,\\
  \label{eq:b}
  2 b&=&k(k-1)\Lambda_2-2 k^2 \Lambda_1~,\\
  \label{eq:c}
  6 c&=&-k(k-1) [(k-2)\Lambda_3-3(k-1) \Lambda_2+3 k \Lambda_1]~.
\end{eqnarray}
Since $R(1)<0$, a necessary condition for the normal form $R_{\rm{nf}}(p)$ to capture this behaviour is that  $c>0$, so that $R_{\rm{nf}}(p=1;a=b=0) < 0$.
If $c>0$ and the parameters $a$ and $b$ vary, the shape of $R_{\rm{nf}}(p)$ evolves in a similar way to that of $R(p)$ when parameters $\alpha$ and $\beta$ vary. 
This is illustrated by Fig.~\ref{fig:Normal_Form_NF_R}(b) which shows an example in which both $R(p)$ and $R_{\rm{nf}}(q)$ (cf. thin and bold lines) exhibit the expected shapes for $\beta<\beta_{\rm{SNT}}$, i.e. in regime III.

If the condition $c>0$ is satisfied, the fixed points of the minimal model (i.e. the zeros of $R_{\rm{nf}}$) reproduce the three dynamical regimes observed for the full model (see Fig.~\ref{fig:Normal_Form_NF_R}(a)). Regime I (non-active) is bounded from above by a line $a=0$ (for $b<0$) of TC bifurcations  and by a parabolic line $a=-b^2/4c$ (for $b<0)$ of SN bifurcations.  Regime II (active) is observed for any $a>0$, i.e. above the TC line which exists for any value of $b$. Regime III (bi-stable) is bounded by the TC bifurcation line from above and by the SN bifurcation line from below (i.e. it exists provided $a<0$ and $b>2\sqrt{-ca}$).  All the three regimes meet at the co-dimension two SNT crossing bifurcation point located at $a=b=0$. 

In fact, the minimal model with $c>0$ exhibits all three regimes on the plane $(\alpha,\beta)$ corresponding to the full model.  In order to prove this, it is necessary to show that the domain in the  $(\alpha,\beta)$ parameter space where $c>0$ (with $c$ given by Eq.~\eref{eq:c}) covers a finite neighbourhood of the SNT crossing bifurcation point where the three regimes meet. In Appendix~\ref{sec:Validity_NF}, we prove this for c-time dynamics. A rigorous proof for d-time dynamics is more challenging but numerical analyses suggest that this also holds for d-time dynamics on random regular graphs with any $k$.

The minimal model leads to several interesting conclusions. First, it shows that the co-dimension of the SNT crossing bifurcation is 2. Second, it demonstrates that considering synergistic effects associated with up to 3 neighbours is sufficient to observe three regimes for synergistic invasions on random regular graphs with any degree $k$. This follows from the fact that the coefficients $a$, $b$ and $c$ depend only on the transmission rates $\Lambda_1$, $\Lambda_2$ and $\Lambda_3$. Synergistic effects associated with more than 3 neighbours may play a role on the details of invasions but do not affect their qualitative behaviour.

\section{Numerical simulations}
\label{sec:numerics}
The above analysis performed within the single-site mean-field approximation is inherently inexact  and thus needs to be validated by exact numerical simulations. For bifurcation analysis, it appears convenient to calculate numerically the rate function $R(p)$ entering Eq.~\eref{eq:markov} and investigate the behaviour of its roots with variation of parameters of the model. 
The advantage in analysing numerically the rate function rather than just calculating the time series for concentration of infected nodes
is the following.
It gives unambiguous criteria for location of (i) the SN bifurcations, i.e. discontinuous transitions in concentration of infected nodes, and (ii) unstable equilibrium points (cf.~\cite{Oliveira_2015:PRE,Chae_2015:New_J_Phys,Varghese_2013:PRE}).
The rate function can be used for location of the TC bifurcations as well although the procedure for finding critical points for continuous transitions is well established~\cite{Ferreira_2011:PRE,Ferreira_2012:PRE}.

\begin{figure}
  \centering
  \includegraphics[clip=true,width=0.45\textwidth]{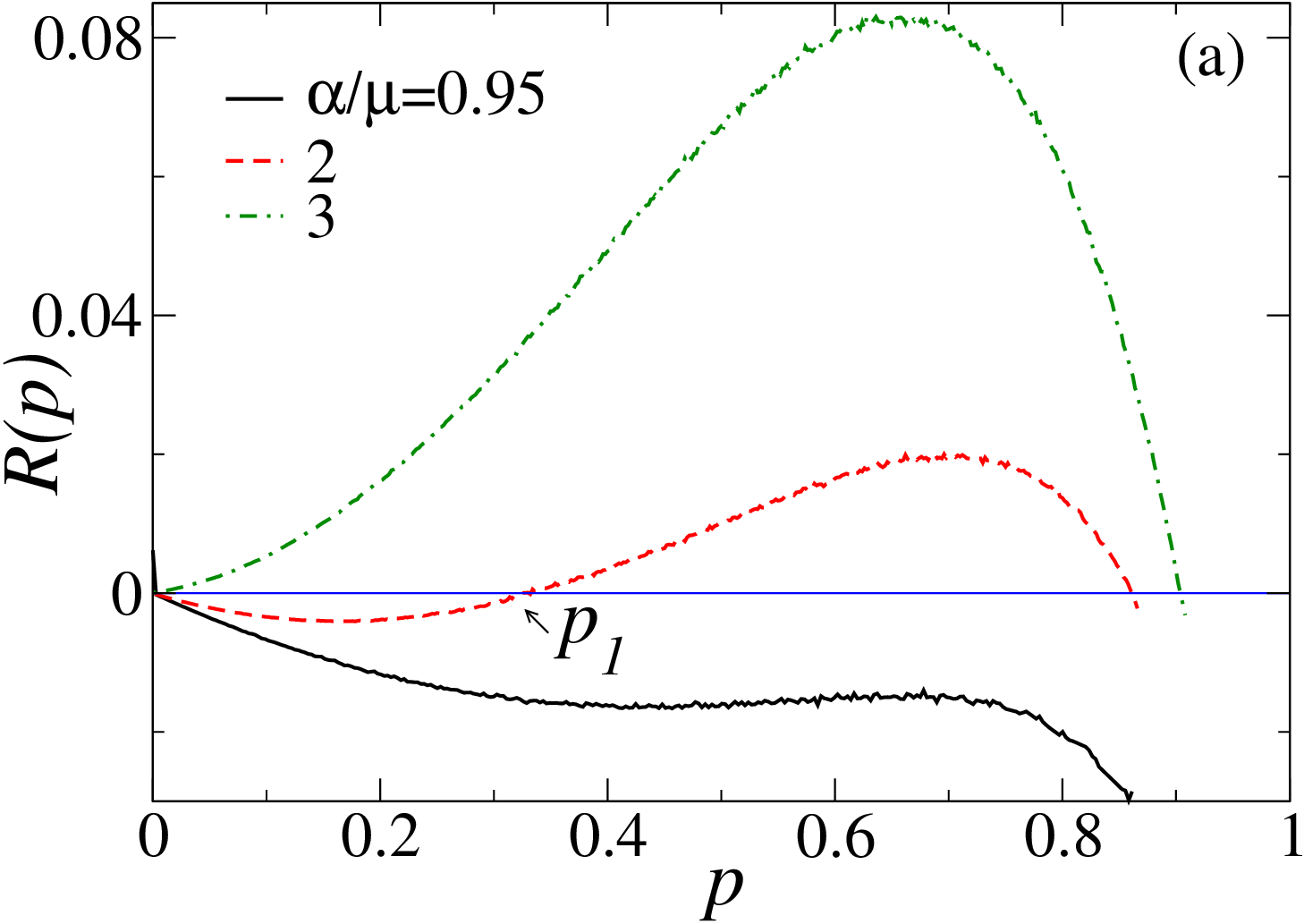}
\quad
\includegraphics[clip=true,width=0.45\textwidth]{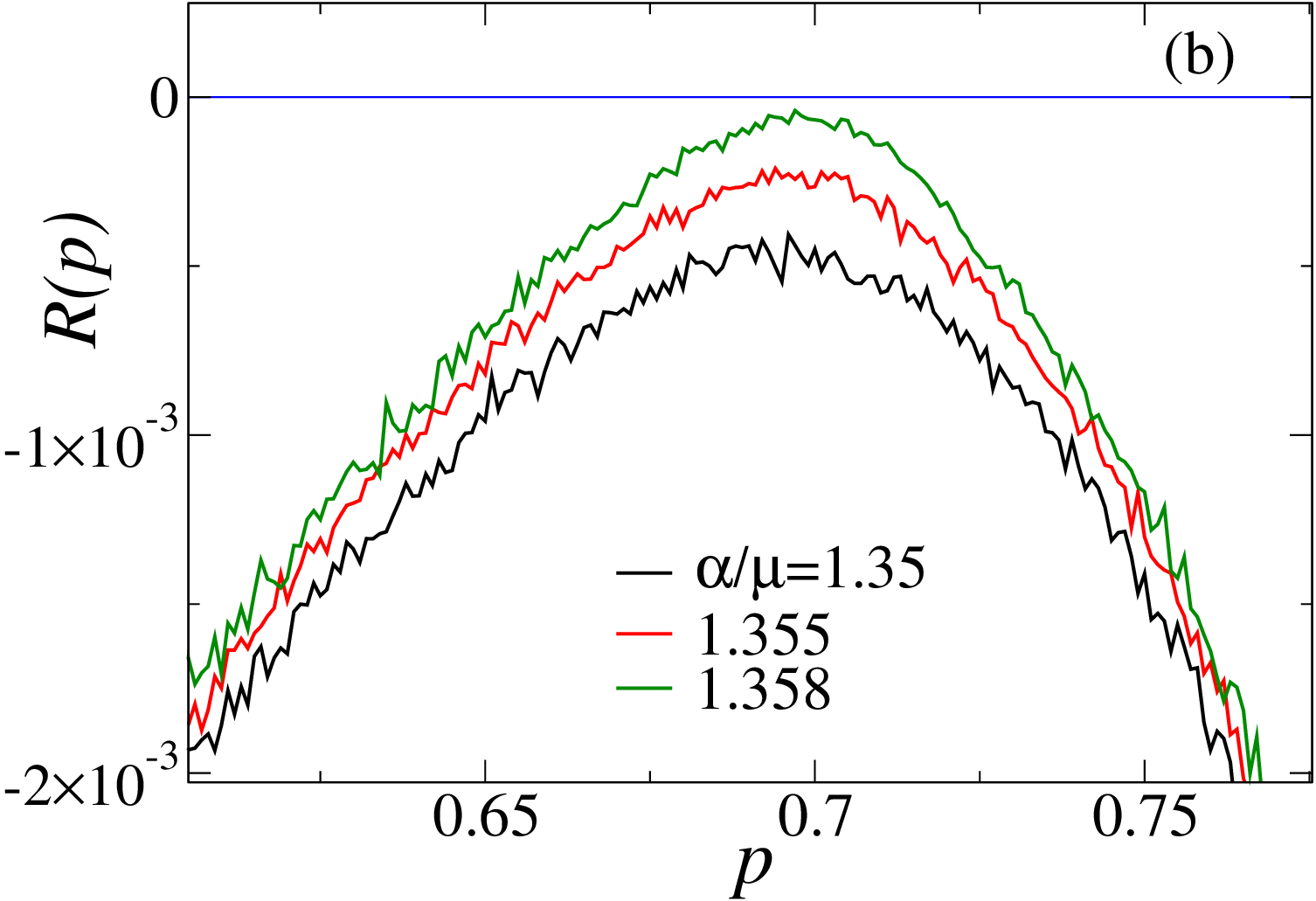}
\caption{\label{fig:numerics}
  The dependence of the rate function, $R(p)$, calculated numerically on concentration of infected nodes, $p$, for d-time SIS process exhibiting S-synergy in transmission on random 10-regular  graphs ($N=10^5$) with $\mu=0.1$, $\beta=-0.4$ and different values of  $\tilde{\alpha}=\alpha/\mu$ as indicated in the legends.
  The panel {\bf (a)} shows $R(p)$ for non-active (solid line, $p(0)=1$), bi-stable (dashed, $p(0)=0.307$) and active (dot-dashed, $p(0)=10^{-3}$) regimes.
  The panel {\bf (b)} displays  $R(p)$, for values of $\tilde{\alpha}$ near the SN bifurcation ($p(0)=1$ for all the curves). 		
	}
\end{figure}

The shape of the rate function calculated numerically follows that predicted by single-site mean-field analysis (cf. the curves shown in Fig.~\ref{fig:phase_diag_KR_10_S}(c) and Fig.~\ref{fig:numerics}(a) shown for SIS processes exhibiting S-synergy).
As follows from Fig.~\ref{fig:numerics}(a), the roots of the rate function (and thus the concentration of infected nodes in stable and unstable equilibrium states) can be identified.
This is particularly important for the roots corresponding to unstable equilibria (see the middle root at $p_1\simeq 0.33$ for the dashed curve in Fig.~\ref{fig:numerics}(a)).
The range of $p$ for which the rate function in bi-stable regime is available numerically significantly depends on initial conditions, i.e. on $p(0)$.
If $p(0) \simeq p_1$, then the SIS process can either go to extinction or invasion and $R(p)$ becomes available for practically the whole range of concentration of infected nodes. 
In particular, the dashed curve shown in Fig.~\ref{fig:numerics}(a) is calculated for $p(0)=0.307$ averaged over $10^4$ network configurations  resulting in $p_1\simeq 0.339 \pm 0.007$. 

The location of the SN bifurcation can be estimated by finding the values of the rate function at the maximum for different values of  $\tilde{\alpha}$ and extrapolating them to zero.
An example of evolution of $R(p)$ near the SN bifurcation is shown in Fig.~\ref{fig:numerics}(b) leading to an estimate for $\tilde{\alpha}_{\rm{SN}}\simeq 1.3595   \pm 0.0005$. 

Overall, we found that numerical analysis supports qualitatively  all the main findings within a simple single-site mean-field approximation.
As expected, the quantitative agreement between MC and mean-field data are not perfect.
Indeed, comparing  Figs.~\ref{fig:phase_diag_KR_10_S}(c) and Fig.~\ref{fig:numerics}(a), we can see that the evolution of the shape of the rate function with $\tilde{\alpha}$ is similar for both mean-field and MC  data but the values of $R(p)$ for the same $\tilde{\alpha}$ are quite different in both panels (cf. e.g. solid lines for $\tilde{\alpha}=0.95$). 
Consequently, the open symbols representing MC data in Figs~\ref{fig:phase_diag_KR_10_S}(a) and ~\ref{fig:phase_diag_KR_10_I}(a) deviate from the continuous bifurcation lines obtained analytically.
The disagreement between MC  and mean-field data are sufficiently significant in the location of SN bifurcations for very negative values of $\beta$ in the case of S-synergy (cf. location of squares for MC data and dashed lines for mean-field in Fig.~\ref{fig:phase_diag_KR_10_S}(a) and, especially, in  Fig.~\ref{fig:phase_diag_KR_3_S}(a) for $3$-regular graph). 

This is not surprising because the single-site mean-field approximation completely ignores the dynamical correlations which can be quite important for synergistic SIS processes.
However, the main qualitative features such as existence of three different regimes are well reproduced by single-site mean-field analysis.

  Several approaches have been developed for dealing with dynamical correlations in synergy-free spreading processes~\cite{Ferreira_2013:EurPhysJ,Gleeson_2013:PRX,Boguna_2013:PRL,Mata_2014:NewJPhys,Cai_2014:PRE,Luo_2014:Physica,Cai_2016:PRL}.
  The synergy effects bring new features in the dynamics which do not permit to use straightforwardly the results obtained for synergy-free cases.
  A possible way forward could consist in using a 
  two-site approximation for $ P(C_n)$ in Eq.~\eref{eq:rate-general-2} which we hope to address  in future analysis.

\section{Conclusions}

To conclude, we presented the single-site mean-field analysis for synergistic SIS processes spreading  on random k-regular graphs.
  The synergy effects investigated account for possible non-linear cooperative effects in transmission of infection between nodes in a network.
  In particular, two cases were investigated in which the individual rate of transmission  from an infected node to a susceptible one depends on number of neighbours of susceptible node being either in susceptible (S-synergy) or infected (I-synergy) states.
  The synergistic transmission is parameterised by introducing two parameters, the inherent transmission rate (i.e. the rate in the synergy-free limit) and the strength of synergy, in such a way that the synergistic transmission rates are continuous functions of both of these parameters and also of discrete number of infected neighbours affecting the transmission.
  The latter property distinguishes the synergy model  from the popular threshold models.
  Moreover, the continuity of the synergistic rates in all variables makes possible to analyse  phase diagrams  in two-parameter space and reveal quite a rich picture even in the simplest topological case of k-regular graphs.
  In particular, the phase diagrams for synergistic SIS process on k-regular graphs (with $k>2$) exhibit three regions found both numerically and analytically:
  non-active, active and bi-stable.
  These regions are separated by lines of transcritical and saddle-node bifurcations which cross (interact) at the saddle-node-transcritical bifurcation.
  The latter bifurcation point, to our knowledge, has not been observed within the threshold models. 

Also, we developed a numerical procedure based on analysis of the rate function for detection of the saddle-node bifurcations, transcritical bifurcations at the boundary between stable and active regimes and unstable equilibria in bi-stable regime. In these cases, the standard analysis of the time-series for concentration of infected nodes is not very helpful. In contrast, analysis of the rate of change in concentration of infected nodes appears to be sufficiently revealing.

      The synergy model is general and can be applied to various spreading processes (e.g. SI, SIS, SIR, contact, catalytic reaction-diffusion and others)  on networks of different topology.
      Our preliminary analysis of synergistic processes on other networks (Erd\"os-R\'enyi, binary and scale-free) shows even more complex bifurcation diagrams with possibility of appearance of several SNT, SN, TC and cusp bifurcations. 
      Currently, the synergy effects are analysed only for transmission rates but they can be straightforwardly incorporated for recovery rates as well.
      Analytically, the main challenge remains in accurate description of the dynamical correlations for synergistic processes, which we hope to address in the future.

\ack{Acknowledgements}
FJPR acknowledges financial support from the Carnegie Trust.

\appendix

\section{Stability of fixed points}
\label{app:stability}
In this section, we study the stability of fixed points of the solutions of the proposed models in the single-site mean-field approximation which obey Eq.~\eref{eq:markov} with the rate function $R(p)$ given by Eq.~\eref{eq:markov_mean_1}. Stability of fixed points can be qualitatively understood from the graphical representation of $R(p)$ but here we present a more rigorous analysis based on the Lyapunov function method~\cite{Bof2018,Arnold_Book-ODEs}. 

Before dealing with the synergistic epidemic model, consider a generic time-invariant dynamical system described by a variable $x \in \mathbb{R}$ which obeys d-time dynamics given by the difference equation,
\begin{equation}
\label{eq:x_dynamics}
 x(t+\delta t) =x(t)+f(x(t)) \delta t~. 
\end{equation}
Assume that the real-valued function $f(x)$ satisfies 
\begin{equation}
\label{eq:fx}
       f(x)=\cases{>0& for $x \in (a,x^*)$\\
=0& for $x=x^*$\\
<0& for $x \in (x^*,b)$~,\\}
\end{equation}
in some interval $(a,b)$ of $x$, where $a$ and $b$ are real parameters, $a<b$. 

\begin{theorem}
\label{th:dtime-xstable}
 The point $x=x^*$ is an asymptotically stable fixed point in the interval $(a,b)$.
\end{theorem}

\noindent\emph{Proof.} The function
\begin{equation}
 V(x)=-\int_{x^*}^x f(y) dy
\end{equation}
is positive definite for any value of $x\in(a,b)$ except at $x=x^*$ where it is  $V(x^*)=0$. 

In addition, the variation of $V(x)$ with time along any trajectory $x(t)$ of the system,
\begin{equation}
 \frac{\delta V}{\delta t} = \frac{V(x(t+\delta t))-V(x(t))}{\delta t}=-\frac{1}{\delta t}\int_{x(t)}^{x(t)+f\delta(t)} f(y) dy,
\end{equation}
is negative everywhere in $(a,b)$ except at $x=x^*$. 

Indeed, for $x \in (x^*,b)$, $f(x)<0$ and one obtains
\begin{equation}
 \delta V = - \int_{x(t)}^{x(t)+|f(x(t))|\delta t} |f(y)| dy < 0~.
\end{equation}

Similarly, for $x \in (a,x^*)$, $f(x)>0$ and one obtains
\begin{equation}
 \delta V = - \int_{x(t)-|f(x(t))|\delta t}^{x(t)} |f(y)| dy < 0~.
\end{equation}

This proves that $V(x)$ is a Lyapunov function and $x=x^*$ is an asymptotically stable fixed point in the interval $(a,b)$. $\square$

\begin{corollary}
\label{cor:ctime-xstable}
In the c-time limit (i.e. for infinitesimal $\delta t \rightarrow dt$), the system \eref{eq:x_dynamics} has a fixed point at $x=x^*$ which is asymptotically stable in $(a,b)$.
\end{corollary}

\emph{Proof.} In the c-time limit, the difference equation \eref{eq:x_dynamics} reduces to $dx/dt=f(x)$ and the variation of $V(x)$ with respect to time along a trajectory $x(t)$ of the system is
\begin{equation}
 \frac{d V}{dt} = -(f(x(t)))^2 < 0
\end{equation}
for any $x \in (a,b)-\{x^*\}$.  Therefore, $V(x)$ is a Lyapunov function and this proves the corollary. $\square$

We now use these results to analyse the stability of fixed points for the concentration of infected nodes, $p(t)$, given by Eq.~\eref{eq:markov} with $R(q)$ given by Eq.~\eref{eq:markov_mean_1}. The results apply in general to both S- and I-synergy models with both d- and c-time dynamics.

\begin{corollary}[Stability in regime I]
\label{co:RegimeI}
 Consider regime I which is characterised by a single fixed point at $p=p_0=0$. This fixed point is globally asymptotically stable in the feasible interval of $p \in (0,1]$. 
\end{corollary}

\emph{Proof.} Since $p_0=0$ is the only fixed point and the infection rate satisfies $R(0)=0$ and $R(1) < 0$, it is clear that $R(p)<0$ for any $p\in (0,1]$.  For c-time dynamics, this system is equivalent to the system given by Eq.~\eref{eq:x_dynamics} with $x\in (x^*,b)$. Accordingly, $p_0=0$ is globally asymptotically stable in $(0,1]$ by Theorem \ref{th:dtime-xstable}. Similarly, stability in the d-time case follows from corollary \ref{cor:ctime-xstable}. $\square$

\begin{corollary}[Stability in regime II]
\label{co:RegimeII}
 In regime II, there are the non-active ($p=p_1>0$) and active ($p=p_0=0$)  fixed points. The active fixed point, $p_1>0$, is globally asymptotically stable in the feasible interval $(0,1]$ and the non-active fixed point $p_0=0$ is unstable. 
\end{corollary}

\emph{Proof.} Let us first consider the fixed point for the active regime. The conditions $R(0)=0$ and $R(1) < 0$ imply that $R(p)>0$ for $p\in (0,p_1)$ and $R(p)<0$ for $p\in (p_1,1]$. Therefore, the problem reduces to that of the system \ref{eq:x_dynamics} with $a=0$, $x^*=p_1$ and $b=1$. By Theorem \ref{th:dtime-xstable} and Corollary \ref{cor:ctime-xstable} it is clear that the fixed point at $p=p_1$ is globally asymptotically stable in the feasible interval $(0,1]$ for both d- and c-time dynamics. The fixed point at $p_0=0$ corresponding to the non-active regime is therefore unstable.  $\square$

\begin{corollary}[Stability in regime III]
\label{co:RegimeIII}
 In regime III, there are three fixed points: the non-active fixed point at $p=p_0=0$, the active fixed point at $p=p_2>0$ and a fixed point with intermediate $p=p_1 \in (p_0,p_2)$. Stability of these points is as follows for both d- and c-time dynamics:
 \begin{enumerate}
\item The fixed point at $p=p_0=0$ is locally asymptotically stable in the interval $(0,p_1)$.
\item The fixed point at $p=p_1$ is unstable.
\item The fixed point at $p=p_2$ is locally asymptotically stable in the interval $(p_1,1]$.
\end{enumerate}
 \end{corollary}

 \emph{Proof.} Local stability of $p=p_0=0$ and $p=p_2$ follows from Theorem \ref{th:dtime-xstable} and Corollary \ref{cor:ctime-xstable} using the conditions $R(0)=0$ and $R(1) < 0$ which imply the following behaviour for $R(p)$:
\begin{equation}
\label{eq:fx1}
       R(p)=\cases{<0& for $p \in (0,p_1)$\\
>0& for $p \in (p_1,p_2)$\\
<0& for $p \in (p_2,1]$~.\\}
\end{equation}
Since the basins of attraction of the fixed points $p=p_0=0$ and $p=p_2$ cover the whole feasible interval $(0,1]$ except for the point $p=p_1$, we conclude that the fixed point at $p=p_1$ is unstable.

\section{Validity of the minimal model}
\label{sec:Validity_NF}
In this appendix, we show for c-time that the condition $c>0$ (see Eq.~\eref{eq:c}) defines a region in the $(\alpha,\beta)$ parameter space which contains the SNT crossing bifurcation for both S- and I-synergy if $k>2$.

The transmission rate for c-time dynamics reduces to $\Lambda_n=n \lambda_n$ and this allows the condition $c>0$ to be expressed as follows:
\begin{equation}
(k-2)\lambda_3-2(k-1)\lambda_2+k\lambda_1<0~.
\end{equation}
This condition is satisfied for any value of $\alpha$ and $\beta \in B$, where 
\begin{equation}
 B=\cases{\left(-\rm{ln}\left(\frac{k}{k-2}\right),0 \right)& for S-synergy\\
 \left(0,\rm{ln}\left(\frac{k}{k-2}\right) \right) & for I-synergy\\}
\end{equation}
From Eqs.~\eref{eq:snt_beta} and \eref{eq:snt_beta_I} it is clear that the SNT crossing bifurcation point belongs to $B$ for both S- and I-synergy.

\section*{References}

\begin{thebibliography}{10}
  \providecommand{\url}[1]{{#1}}
\providecommand{\urlprefix}{URL }
\expandafter\ifx\csname urlstyle\endcsname\relax
  \providecommand{\doi}[1]{DOI~\discretionary{}{}{}#1}\else
  \providecommand{\doi}{DOI~\discretionary{}{}{}\begingroup
  \urlstyle{rm}\Url}\fi

\bibitem{Arnold_Book-ODEs}
Arnol'd, V.I.: {Ordinary Differential Equations}.
\newblock Springer-Verlag Berlin Heidelberg, Heidelberg (1992)

\bibitem{Barrat_08:book}
Barrat, A., Barth\'{e}lemy, M., Vespignani, A.: Dynamical Processes on Complex
  Networks.
\newblock Cambridge University Press, Cambridge (2008)

\bibitem{Baxter_2010:PRE}
Baxter, G.J., Dorogovtsev, S.N., Goltsev, A.V., Mendes, J.F.F.: Bootstrap
  percolation on complex networks.
\newblock Phys. Rev. E \textbf{82}, 011103 (2010)

\bibitem{Bof2018}
Bof, N., Carli, R., Schenato, L.: {Lyapunov Theory for Discrete Time Systems}.
\newblock Tech. rep., Universita degli studi di Padova (2018).
\newblock \urlprefix\url{http://arxiv.org/abs/1809.05289}

\bibitem{Boguna_2013:PRL}
Bogu\~n\'a, M., Castellano, C., Pastor-Satorras, R.: 
Nature of the Epidemic Threshold for the Susceptible-Infected-Susceptible Dynamics in Networks. 
\newblock Phys. Rev. Lett. \textbf{111}, 068701 (2013)

\bibitem{Bottcher_2017:Sci_Rep}
B\"ottcher, L., Lukovic, M., Nagler, J., Havlin, S., Herrmann, H.J.: Bootstrap
  percolation on complex networks.
\newblock Sci. Rep. \textbf{7}, 41729 (2017)

\bibitem{Bottcher_2017:PRL}
B\"ottcher,  M., Nagler, S., Herrmann, H.J.: Critical Behaviors in Contagion Dynamics. 
\newblock Phys. Rev. Lett. \textbf{118}, 088301  (2017)

\bibitem{Bollobas2001}
Bollob\'{a}s, B.: Random Graphs. 
       \newblock Cambridge University Press, Cambridge (2001)
       
\bibitem{Chae_2015:New_J_Phys}
Chae, H., Yook, S.H., Kim, Y.: Discontinuous phase transition in a core contact
  process on complex networks.
\newblock New J. of Physics \textbf{17}, 023039 (2015)

\bibitem{Cai_2014:PRE}
    Cai, C.-R.,   Wu, Z.-X., Guan, J.-Y.: 
Effective degree Markov-chain approach for discrete-time epidemic processes on uncorrelated networks. 
\newblock Phys. Rev. E \textbf{90}, 052803 (2014)

\bibitem{Cai_2016:PRL}
  Cai, C.-R.,   Wu, Z.-X., Chen, M. Z. Q., Holme, P., Guan, J.-Y.: 
Solving the Dynamic Correlation Problem of the Susceptible-Infected-Susceptible Model on Networks. 
\newblock Phys. Rev. Lett. \textbf{116}, 258301 (2016)

\bibitem{Centola_2010:Science}
  Centola, D.:
  The Spread of Behavior in an Online Social Network Experiment. 
\newblock Science \textbf{329}, 1194 (2010)

\bibitem{Dodds_2004:PRL}
Dodds, P.S., Watts, D.J.: Universal behavior in a generalized model of
  contagion.
\newblock Phys. Rev. Lett. \textbf{92}, 218701 (2004)

\bibitem{Dorogovtsev_2008:RevModPhys}
  Dorogovtsev, S. N., Goltsev, A. V.: Critical phenomena in complex networks.
 \newblock  Rev. Mod. Phys. \textbf{80}, 1275–1335 (2008).

\bibitem{VandenDriessche2002}
van~den Driessche, P., Watmough, J.: {Reproduction numbers and sub-threshold
  endemic equilibria for compartmental models of disease transmission}.
\newblock Math. Biosci. \textbf{180}(1-2), 29--48 (2002)

 \bibitem{Ferreira_2012:PRE}
Ferreira, S.C., Castellano, C., Pastor-Satorras, R.: Epidemic thresholds of the
  susceptible-infected-susceptible model on networks: A comparison of numerical
  and theoretical results.
\newblock Phys. Rev. E \textbf{86}, 041125 (2012)

\bibitem{Ferreira_2013:EurPhysJ}
Ferreira, R.S., Ferreira, S.C.: Critical behavior of the contact process on small-world networks. 
\newblock Eur. Phys. J. B \textbf{86},  462 (2013) 

\bibitem{Ferreira_2011:PRE}
Ferreira, S.C., Ferreira, R.S., Castellano, C., Pastor-Satorras, R.:
  Quasistationary simulations of the contact process on quenched networks.
\newblock Phys. Rev. E \textbf{84}, 066102 (2011)

\bibitem{Gleeson_2012:PRE}
Gleeson, J.P., Melnik, S., Ward,  J. A., Porter, M. A., Mucha,  P. J.: Accuracy of Mean-Field Theory for Dynamics on
Real-World Networks, Phys. Rev. E \textbf{85}, 026106 (2012).

\bibitem{Gleeson_2013:PRX}
Gleeson, J.P.: Binary-state dynamics on complex networks: Pair approximation
  and beyond.
\newblock Phys. Rev. X \textbf{3}, 021004 (2013)

\bibitem{Gomez-Gardenes_Lotero_Taraskin_FJPR2015}
  Gomez-Gardenes, J., Lotero, L., Taraskin, S., Perez-Reche, F.:
Explosive Contagion in Networks. 
\newblock Sci. Rep. \textbf{6}, 19767 (2015)

\bibitem{Gross_2006:PRL}
Gross, T., D'Lima, C.J.D., Blasius, B.: Epidemic dynamics on an adaptive
  network.
\newblock Phys. Rev. Lett. \textbf{96}, 208701 (2006)

\bibitem{Gubbins-Gilligan_2000}
Gubbins, S., Gilligan, C.A., Kleczkowski, A.: {Population dynamics of
  plant-parasite interactions: Thresholds for invasion}.
\newblock Theor. Popul. Biol. \textbf{57}, 2000 (2000)

\bibitem{Guo_2013:PRE}
    Guo, D., Trajanovski, S., van de Bovenkamp, R., Wang, H., Van Mieghem, P.:
    Epidemic threshold and topological structure of susceptible-infectious-susceptible
    epidemics in adaptive networks.
    \newblock Phys. Rev. E \textbf{88}, 042802 (2013)
    
\bibitem{HethcoteSIAM2000}
Hethcote, H.W.: {The Mathematics of Infectious Diseases}.
\newblock SIAM Rev. \textbf{42}(4), 599 (2000)

\bibitem{Juul_2018_synergy:Chaos}
  Juul, J.S.,  Porter, M.A.: Synergistic effects in threshold models on networks.
  \newblock Chaos \textbf{28}, 013115 (2018)

\bibitem{Liu_2007:PRL}
  Liu, D.-J., Guo, X., Evans, J.W.: 
Quadratic Contact Process: Phase Separation
with Interface-Orientation-Dependent Equistability
\newblock Phys. Rev. Lett. \textbf{98}, 050601 (2007)

\bibitem{Liu_ChengLai_PRE2017}
Liu, Q.H., Wang, W., Tang, M., Zhou, T., Lai, Y.C.: Explosive spreading on
  complex networks: The role of synergy.
\newblock Phys. Rev. E \textbf{95}, 042320 (2017)

\bibitem{Liu1987}
Liu, W.m., Hethcote, H.W., Levin, S.A.: {Dynamical behavior of epidemiological
  models with nonlinear incidence rates}.
\newblock J. Math. Biol. \textbf{25}(4), 359--380 (1987)

\bibitem{Ludlam_2011:Interface}
  Ludlam, J.J, Gibson, G. J., Otten, W.,   Gilligan, C.A.:
Applications of percolation theory to fungal spread with synergy. 
 \newblock  J. R. Soc. Interf. \textbf{9}, 949 (2011).

\bibitem{Luo_2014:Physica}
 Luo, X.F., Zhang, X.G., Sun, G.Q., Jin, Z.: Epidemical dynamics of SIS pair ap-
 proximation models on regular and radnom networks.
 \newblock    Physica A \textbf{410}, 144 (2014)

\bibitem{Marceau_2010:PRE}
Marceau, V., No\"el, P.A., H\'ebert-Dufresne, L., Allard, A., Dub\'e, L.J.:
  Adaptive networks: Coevolution of disease and topology.
\newblock Phys. Rev. E \textbf{82}, 036116 (2010)

\bibitem{Majdandzic_2014:NaturePhys} 
Majdandzic, A., Podobnik, B., Buldyrev, S. V.,  Kenett, D. Y., Havlin, S., Stanley, E. H.:
Spontaneous recovery in dynamical networks. 
\newblock  Nature Phys. \textbf{10}, 34 (2014)

\bibitem{Mata_2014:NewJPhys}
  Mata, S.A., Ferreira, R.S.,   Ferreira, S.C.:
   Heterogeneous pair-approximation for the contact
process on complex networks. 
\newblock   New J. Phys. \textbf{16}, 053006 (2014)
  
\bibitem{Min_2018:Sci_Rep}
Min, B., Miguel, M.S.: Competing contagion processes: Complex contagion
  triggered by simple contagion.
\newblock Sci. Rep. \textbf{8}, 10422 (2018)


\bibitem{Oliveira_2015:PRE}
de~Oliveira, M.M., da~Luz, M.G.E., Fiore, C.E.: Generic finite size scaling for
  discontinuous nonequilibrium phase transitions into absorbing states.
\newblock Phys. Rev. E \textbf{92}, 062126 (2015)

\bibitem{Pastor-Satorras_15:review}
Pastor-Satorras, R., Castellano, C., Van~Mieghem, P., Vespignani, A.: Epidemic
  processes in complex networks.
\newblock Rev. Mod. Phys. \textbf{87}, 925 (2015)

\bibitem{Perez_Reche_2011:PRL}
P\'erez-Reche, F.J., Ludlam, J.J., Taraskin, S.N., Gilligan, C.A.: Synergy in
  spreading processes: From exploitative to explorative foraging strategies.
\newblock Phys. Rev. Lett. \textbf{106}, 218701 (2011)

\bibitem{Pianegonda_Fiore_2015:JStatMech}
Pianegonda, S., Fiore, C.E.: Effect of diffusion in simple discontinuous
absorbing transition models. 
\newblock J. Stat. Mech. P08018 (2015) 

\bibitem{Porter-Gleeson_Book2016}
Porter, M. A., Gleeson, J. P.: Dynamical Systems on Networks. 
\newblock Springer International Publishing, Cham (2016)

\bibitem{Schlogl_1972}
  Schl{\"o}gl, F.: Chemical reaction models for non-equilibrium phase transitions. 
\newblock  Z. Phys. \textbf{253}, 147 (1972)

\bibitem{Saputra2010}
Saputra, K., Veen, L., Quispel, G.: The saddle-node-transcritical bifurcation in a population model with constant rate harvesting. 
\newblock
Discret. Contin. Dyn. Syst. - Ser. B \textbf{14}, 233 (2010)

\bibitem{Silva_Oliveira_2011:JPhysA}
da~Silva, E. F., de~Oliveira, M.M.: Critical discontinuous phase transition in the
threshold contact process. 
\newblock J. Phys. A: Math. Theor. \textbf{44},  135002 (2011) 

\bibitem{Strogatz_NonlinearBook}
Strogatz, S.: Nonlinear Dynamics and Chaos.
\newblock Addison-Wesley, Reading, Massachusetts (1994)

\bibitem{Taraskin-PerezReche_PRE2013_Synergy}
Taraskin, S.N., P\'erez-Reche, F.J.: Effects of variable-state neighborhoods
  for spreading synergystic processes on lattices.
\newblock Phys. Rev. E \textbf{88}, 062815 (2013)

\bibitem{Watts_2002:PNAS}
  Watts, D. J.: A simple model of global cascades on random networks.
\newblock   Proc. Natl Acad. Sci. USA  \textbf{99}, 5766–5771 (2002)
    
\bibitem{Varghese_2013:PRE}
Varghese, C., Durrett, R.: Phase transitions in the quadratic contact process
  on complex networks.
\newblock Phys. Rev. E \textbf{87}, 062819 (2013)

\end{thebibliography}

\end{document}